\documentclass{SciPost}
\hypersetup{
    colorlinks,
    linkcolor={red!50!black},
    citecolor={blue!50!black},
    urlcolor={blue!80!black}
}

\usepackage[bitstream-charter]{mathdesign}
\DeclareSymbolFont{usualmathcal}{OMS}{cmsy}{m}{n}
\DeclareSymbolFontAlphabet{\mathcal}{usualmathcal}

\fancypagestyle{SPstyle}{
\fancyhf{}
\lhead{\colorbox{scipostblue}{\bf \color{white} ~SciPost Physics Core }}
\rhead{{\bf \color{scipostdeepblue} ~Submission }}

\fancyfoot[C]{\textbf{\thepage}}
}

\usepackage[normalem]{ulem}
\usepackage{graphicx}
\usepackage{amsmath}
\usepackage[T1]{fontenc}
\usepackage{calligra}
\usepackage{array}
\usepackage{xcolor}
\usepackage{float}
\usepackage{soul}
\usepackage{bm}
\usepackage[toc,page]{appendix}
\usepackage{braket}
\usepackage{wrapfig}
\usepackage[english]{babel}
\DeclareUnicodeCharacter{2212}{\ensuremath{-}}
\hypersetup{colorlinks,linkcolor=blue,citecolor=blue,urlcolor=blue}

\newcommand{\fo}[1]{\left(#1\right)}
                 \newcommand{\fob}[1]{\left[#1\right]}

                \newcommand{\posvector}{\mathbf{p}}
\newcommand{\rottot}{\theta}
\newcommand{\gra}{G}                                    \newcommand{\tmd}{TMD}
\newcommand{\dos}{DoS}
\newcommand{\rotlayg}{\rottot_\gra}                     \newcommand{\rotlayt}{\rottot_{\tmd}}
                                  \newcommand{\oig}{\Delta}

\newcommand{\tsm}{M}                                    \newcommand{\tsx}{X}
\newcommand{\gc}{C}
\newcommand{\acc}{a_{\gc\gc}}
\newcommand{\ag}{a_\gra}                                \newcommand{\at}{a_{\tmd}}
\newcommand{\aone}{\mathbf{a_1}}                        \newcommand{\atwo}{\mathbf{a_2}}
\newcommand{\ux}{\hat{\mathbf{x}}}                      \newcommand{\uy}{\hat{\mathbf{y}}}
\newcommand{\uz}{\hat{\mathbf{z}}}
                                 
\newcommand{\nG}{n_\gra}                                \newcommand{\mg}{m_\gra}
\newcommand{\nt}{n_{\tmd}}                              \newcommand{\mt}{m_{\tmd}}
\newcommand{\Rnm}{\mathbf{R}^{n,m}}
\newcommand{\Rgnm}{\mathbf{R}^{\nG,\mg}_{\gra}}         \newcommand{\Rtnm}{\mathbf{R}^{\nt,\mt}_{\tmd}}
\newcommand{\straintot}{\epsilon}                       \newcommand{\strainmax}{\straintot_{max}}
               
\newcommand{\suba}{A}                                   \newcommand{\subb}{B}
\newcommand{\subab}{\suba/\subb}

\newcommand{\dxx}{d_{\tsx-\tsx}}
\newcommand{\dcm}{d_{\gc-\tsm}}                         \newcommand{\dcx}{d_{\gc-\tsx}}
\newcommand{\tsxu}{\tsx_t}                              \newcommand{\tsxl}{\tsx_b}
\newcommand{\lsv}{\lambda_{SV}}                         \newcommand{\lr}{\lambda_{R}}
\newcommand{\lkm}{\lambda_{KM}}
\newcommand{\prh}{\phi_{R}}                             \newcommand{\lra}{|\lr|}
\newcommand{\cpr}{\cos\prh}                             \newcommand{\spr}{\sin\prh}   
\newcommand{\lpi}{\lambda_{PIA}}                        \newcommand{\gen}{\rho}
\newcommand{\lpid}{\lpi^{\Delta}}
                           
\newcommand{\onsitetb}{\varepsilon}
\newcommand{\ons}{\onsitetb_{\lambda}}
\newcommand{\onsg}{\onsitetb_{\subab}}                  
\newcommand{\onsgz}{\onsg^0}                            \newcommand{\onsgs}{\delta\onsg^{shift}}
\newcommand{\engen}{E}
\newcommand{\bac}{\engen_C}                             \newcommand{\bav}{\engen_V}
\newcommand{\bacz}{\bac^0}                              \newcommand{\bavz}{\bav^0}
\newcommand{\bgt}{\engen^{BG}_{\tmd}}
\newcommand{\oxt}{\onsitetb_{\tsxu}}                      \newcommand{\oxtz}{\oxt^0}
\newcommand{\oxtct}{\delta\oxt^{CT}}
\newcommand{\baf}{f_b}                                  \newcommand{\bafz}{\baf^0}
\newcommand{\dct}{\delta N_{CT}}

\newcommand{\mos}{$\text{MoS}_{2}$}
\newcommand{\mose}{$\text{MoSe}_\text{2}$}
\newcommand{\ws}{$\text{WS}_{2}$}
\newcommand{\wse}{$\text{WSe}_\text{2}$}
\newcommand{\lk}{\textit{Li and Koshino}}
\newcommand{\rashba}{Rashba}
\newcommand{\rashbap}{Rashba phase}

\newcommand{\fithar}{f}                                         \newcommand{\fitharparam}[1]{\fithar_{#1}}
\newcommand{\fitharlsv}{\fitharparam{\lsv}}                     \newcommand{\fitharlra}{\fitharparam{\lr}}
\newcommand{\fitharons}{\fitharparam{\ons}}                     \newcommand{\fitharprh}{\fitharparam{\prh}}
\newcommand{\fitharcpr}{\fitharparam{\cpr}}                     \newcommand{\fitharspr}{\fitharparam{\spr}}

\newcommand{\fitharpar}{\alpha}                                 \newcommand{\fitharparparam}[2]{\fitharpar_{#1}^{#2}}
\newcommand{\fitharparlsv}[1]{\fitharparparam{\lsv}{#1}}        \newcommand{\fitharparlra}[1]{\fitharparparam{\lr}{#1}}
\newcommand{\fitharparons}[1]{\fitharparparam{\ons}{#1}}        
\newcommand{\fitharparcpr}[1]{\fitharparparam{\cpr}{#1}}        \newcommand{\fitharparspr}[1]{\fitharparparam{\spr}{#1}}
        \newcommand{\fitharpargen}[1]{\fitharparparam{\gen}{#1}}

\newcommand{\fitpol}{\mathfrak{f}}                                  \newcommand{\fitpolparam}[2]{\fitpol_{#1,#2}}
\newcommand{\fitpollsv}[1]{\fitpolparam{\lsv}{#1}}              \newcommand{\fitpollra}[1]{\fitpolparam{\lr}{#1}}
\newcommand{\fitpolons}[1]{\fitpolparam{\ons}{#1}}              
\newcommand{\fitpolcpr}[1]{\fitpolparam{\cpr}{#1}}              \newcommand{\fitpolspr}[1]{\fitpolparam{\spr}{#1}}
              \newcommand{\fitpolgen}[1]{\fitpolparam{\gen}{#1}}

\newcommand{\fitpolpar}{\beta}                                  \newcommand{\fitpolparparam}[3]{\fitpolpar_{#1,#2}^{#3}}

    \newcommand{\fitpolpargen}[2]{\fitpolparparam{\gen}{#1}{#2}}

\newcommand{\degre}{^\circ}

\newcommand{\figref}[1]{Figure~\ref{#1}}
\newcommand{\eqnref}[1]{Equation~\ref{#1}}
\newcommand{\tabref}[1]{Table~\ref{#1}}
\newcommand{\apxref}[1]{Appendix~\ref{#1}}
\newcommand{\refcite}[1]{Ref.~\cite{#1}}

\begin{document}

        
    \pagestyle{SPstyle}
    
    \begin{center}{\Large \textbf{\color{scipostdeepblue}{
    Influence of Electrostatic Environment on the Proximity Spin-Orbit Coupling in Graphene on Transition-Metal Dichalcogenides\\
    }}}\end{center}
    
    \begin{center}\textbf{
    Bert Jorissen\textsuperscript{1$\star$},
    Bart Partoens\textsuperscript{1},
    Aires Ferreira\textsuperscript{2} and
    Lucian Covaci\textsuperscript{1}
    }\end{center}
    
    \begin{center}
    {\bf 1} Department of Physics, University of Antwerp, Groenenborgerlaan 171, B-2020 Antwerpen, Belgium
    \\
    {\bf 2} School of Physics, Engineering and Technology and York Centre for Quantum Technologies, University of York, YO10 5DD, York, United Kingdom
    \\[\baselineskip]
    $\star$ \href{mailto:bert.jorissen@uantwerpen.be}{\small bert.jorissen@uantwerpen.be}
    \end{center}

    \section*{\color{scipostdeepblue}{Abstract}}
    \textbf{\boldmath{%
        We investigate the proximity-induced spin-orbit coupling (SOC) in graphene on transition-metal dichalcogenide (\tmd) heterostructures using a tight-binding approach.
        The tight-binding parameters of an effective 4-band model describing the low-energy physics of the graphene layer are extracted as a function of twist angle, material composition (\mos, \mose, \ws, \wse), band alignment, and charge-transfer dipole.
        We investigate the effect of a charge-transfer dipole induced on the chalcogen layer closest to graphene, which is typically neglected in tight-binding models but included in ab initio calculations through charge redistribution.
        In addition, we explore the influence of variations in the band alignment.
        We find that the spin-valley and \rashba\ spin-orbit coupling energies as well as the so-called Rashba phase, governing the orientation of momentum-space spin textures, show a strong dependence on the twist angle and electrostatic tuning.
        Our results suggest that variations in band alignment and charge redistribution contribute to the spread of spin-orbit parameters reported in the literature.
    }}
    
    \vspace{\baselineskip}
    
    \noindent\textcolor{white!90!black}{%
    \fbox{\parbox{0.975\linewidth}{%
    \textcolor{white!40!black}{\begin{tabular}{lr}%
      \begin{minipage}{0.6\textwidth}%
        {\small Copyright attribution to authors. \newline
        This work is a submission to SciPost Physics Core. \newline
        License information to appear upon publication. \newline
        Publication information to appear upon publication.}
      \end{minipage} & \begin{minipage}{0.4\textwidth}
        {\small Received Date \newline Accepted Date \newline Published Date}%
      \end{minipage}
    \end{tabular}}
    }}
    }

    
    \vspace{10pt}
    \noindent\rule{\textwidth}{1pt}
    \tableofcontents
    \noindent\rule{\textwidth}{1pt}
    \vspace{10pt}


    \section{Introduction}
    
        2D material spintronics is a rapidly developing frontier in condensed matter physics that intersects with quantum information, spintronic logic and opto-spintronics~\cite{laird_valleyspin_2013, perkins_ultrafast_2024, gmitra_graphene_2015, ahn_2d_2020, avsar_colloquium_2020, zhao_novel_2025, garcia_spin_2018, sierra_van_2021,perkins_spintronics_2024}.
        Its overarching aim is to exploit the atomically thin nature and unique electronic structure of graphene and related 2D materials to efficiently manipulate the electron's spin.
        A particularly promising direction involves endowing graphene with a sizeable spin-orbit coupling (SOC) as a means to achieve pure electrical control over spin states without the need for ferromagnets.
        While the intrinsic SOC in pristine graphene is extremely weak, of the order of tens of $\mu eV$~\cite{gmitra_band-structure_2009, konschuh_theory_2012, kurpas_spin-orbit_2019, sichau_resonance_2019}, placing it in close contact with a suitable high-SOC substrate or doping it with dilute heavy metal elements has been shown to strongly enhance the spin-orbit effects experienced by low-energy carriers.
        Examples range from hydrogenated graphene~\cite{gmitra_spin-orbit_2013} and graphene decorated with copper ad-atoms~\cite{balakrishnan_giant_2014} to graphene on metallic substrates~\cite{frank_theory_2016, krivenkov_nanostructural_2017} and graphene proximity-coupled to topological insulators~\cite{song_spin_2018}.

        Recently, proximity effects in twisted van der Waals heterostructures of graphene and semiconducting transition-metal dichalcogenides (\tmd) have attracted much attention.
        Such material systems have been fabricated using both monolayer~\cite{fulop_boosting_2021, 
            hoque_all-electrical_2021, 
            rao_ballistic_2023, 
            pierucci_band_2016, 
            kim_band_2015, 
            wang_origin_2016, 
            lu_moire-related_2017, 
            omar_spin_2018, 
            zihlmann_large_2018, 
            yang_strong_2017, 
            wang_strong_2015, 
            ghiasi_large_2017, 
            ghiasi_charge--spin_2019, 
            benitez_tunable_2020, 
            li_gate-tunable_2020, 
            han_tailoring_2026, 
            lan_huang_layer-dependent_2025, 
            ghaebi_tunable_2025, 
            benitez_strongly_2018, 
            yang_twist-angle-tunable_2024, 
            rockinger_tuning_2026, 
            sun_determining_2023 
            }
            and bilayer graphene~\cite{szentpeteri_increasing_2025, wang_quantum_2019, masseroni_spin-orbit_2024, amann_counterintuitive_2022, kedves_stabilizing_2023, asim_high-performance_2025, icking_weak_2026, island_spinorbit-driven_2019, dulisch_electric-field-tunable_2025, tiwari_electric-field-tunable_2021}, in combination with \mos, \mose, \ws\ or \wse.
        Interestingly, these experiments have revealed a twist-angle dependence of the induced SOC~\cite{rockinger_tuning_2026, sun_determining_2023, szentpeteri_increasing_2025, yang_twist-angle-tunable_2024},
        while the effective SOC energy parameters on the graphene layer vary from approximately $0.1$ meV to $10$ meV, which translates into a giant SOC enhancement of up to 3 orders of magnitude compared to non-proximitized graphene. 
        
        A symmetry analysis~\cite{wang_strong_2015, kochan_model_2017, perkins_spintronics_2024} shows that there are three main types of SOC in the low-energy description of graphene-TMD heterostructures, namely intrinsic-like (Kane-Mele), spin-valley (SV) and Rashba (R) SOC.
        The first two, namely Kane-Mele and SV, are spin-conserving SOC terms in the effective graphene-only Hamiltonian, while Rashba SOC strongly mixes spin-up and spin-down states.
        Owing to its intrinsic nature, the Kane-Mele SOC preserves the original point group symmetries of graphene.
        Its main effect is to open a spin-orbit energy gap around the $K$ points~\cite{kane_quantum_2005}.
        In contrast, spin-valley and Rashba SOC originate from broken inversion symmetry and thus lift the spin degeneracy of the energy eigenstates.
        The ensuing momentum-space spin polarisation of the Fermi surface (i.e. the \textit{spin texture}) is determined by the interplay of orbital and SOC terms in graphene's effective Hamiltonian.
        In (untwisted) graphene/TMD with the crystal axes of the two layers aligned (and also for the special case of a $30\degre$ rotation angle between the layers; see later), the Rashba SOC creates a tangential spin winding around the Fermi contours, while SV SOC tilts the spin texture out of the basal $Oxy$ plane.
        The interplay of SV and Rashba SOC thus leads to a non-coplanar spin texture around the $K$ and $K^\prime$ valleys~\cite{perkins_spintronics_2024}.
            
        The experiments on the TMD/graphene heterostructure have motivated a series of first-principles (DFT)~\cite{gmitra_trivial_2016, yu_dirac-rashba_2025, wang_strong_2015, ge_ab_2025, gmitra_graphene_2015, zollner_first-principles_2025, gmitra_proximity_2017, naimer_tuning_2024, naimer_twist-angle_2021, zollner_twist-_2023, lee_charge--spin_2022, szalowski_spinorbit_2023} and tight-binding (TB)~\cite{li_twist-angle_2019, peterfalvi_quantum_2022, khatibi_proximity_2022, zollner_bilayer_2021, alsharari_mass_2016, alsharari_topological_2018, singh_proximity-induced_2019, wang_spinorbit_2021, david_induced_2019} studies.
        These works qualitatively reproduce some of the observed trends in the effective SOC parameters inferred from experimental data.
        However, a considerable spread remains within the theoretical predictions and the experimental picture.
        This spread is particularly large among the experimentally inferred SOC parameters, which depend on the measurement technique and on the model assumptions for parameter extraction. 
        For graphene on \wse, for example, experimentally inferred values of the \rashba\ parameter range from approximately $0.14$ to $14$ meV~\cite{rockinger_tuning_2026, szentpeteri_increasing_2025, fulop_boosting_2021, amann_counterintuitive_2022, tiwari_experimental_2022, yang_strong_2017, zihlmann_large_2018, sun_determining_2023, wang_origin_2016, rao_ballistic_2023, wang_quantum_2019}, whereas theoretical predictions are more closely clustered between $0.20$ and $1.22$ meV~\cite{ge_ab_2025, naimer_twist-angle_2021, gmitra_trivial_2016, lee_charge--spin_2022, khatibi_proximity_2022, zollner_twist-_2023, yu_dirac-rashba_2025, wang_highly_2015}.
        Moreover, sample-to-sample variations in the interface quality, specifics of disorder landscape and strain field also contribute to the significant spread of SOC values in reported data.
        In addition, the explicit and implicit approximations in both DFT and TB schemes may lead to discrepancies when undertaking quantitative comparisons between experiment and theory.

        Crucially, a non-trivial twist angle between the TMD and graphene layers, $\rottot\neq 0\degre, 30 \degre,  60\degre$ (modulo $180\degre$) reduces the point-group symmetry from $C_{3v}$ to $C_3$, providing a geometric knob to tune Fermi-surface spin textures~\cite{sun_determining_2023}.
        The in-plane component of the spin polarisation of the low-energy graphene states is characterized by the so-called \rashbap\ $\prh$.
        Symmetry requires $\prh$ to be a multiple of $\pi$ at the twist angles of $0\degre$ or $30\degre$~\cite{veneri_twist_2022}.
        The twist-angle dependence of the \rashbap\ has been predicted theoretically~\cite{peterfalvi_quantum_2022, naimer_twist-angle_2021, naimer_tuning_2024, szalowski_spinorbit_2023, lee_charge--spin_2022, zollner_twist-_2023, li_twist-angle_2019}
        and its impact directly observed on current-induced spin polarisation experiments~\cite{yang_twist-angle-tunable_2024}.
        This twist-angle dependence of the spin texture is relevant to coupled charge-spin transport phenomena, namely the inverse spin galvanic~\cite{gmitra_trivial_2016, ghiasi_charge--spin_2019, lee_charge--spin_2022,veneri_twist_2022,yang_twist-angle-tunable_2024} and spin Hall effects~\cite{perkins_spin_2024, milletari_covariant_2017}, as well as their Onsager reciprocal phenomena (i.e., spin galvanic and inverse spin Hall effects).
        However, the predicted and experimentally inferred values of $\prh$, as well as its twist-angle dependence, differ between studies.
        Previous calculations have demonstrated that $\prh$ is sensitive to a transverse electric field and to the position of the graphene Dirac point relative to the \tmd\ bands~\cite{naimer_twist-angle_2021, peterfalvi_quantum_2022, zollner_twist-_2023, naimer_tuning_2024}.
        However, the separate influences of the interface charge-transfer dipole and the relative band alignment have not yet been systematically investigated.
        
        In this work, we isolate two electrostatic contributions that can affect the proximity-induced SOC: (i) the local interface energy shift associated with a charge-transfer dipole at the interface of the graphene/TMD heterostructure and (ii) the relative band alignment between graphene and the TMD.
        We vary these contributions independently within a tight-binding model in order to distinguish their effects.
        The charge-transfer dipole is represented by a shift of the on-site energies of the chalcogen atoms closest to graphene, whereas the band alignment determines the position of the Dirac point within the \tmd\ band gap.
        
        When graphene is brought into contact with metals, work-function differences cause charge rearrangement in the layers at the interface and may also drive charge transfer between the layers~\cite{giovannetti_doping_2008}.
        Together, these processes produce an interface dipole and an associated potential step, which we refer to as the \textit{charge-transfer dipole}.
        Charge redistribution and transfer have also been reported for graphene/\tmd\ heterostructures~\cite{coy_diaz_interface_2014}, where electrons transfer to the \tmd\ layer, giving graphene weak $p$-type doping~\cite{yun_electronic_2025, fang_van_2022, sukhanova_induced_2020,ma_first-principles_2011, dappe_charge_2020, lan_huang_layer-dependent_2025, coy_diaz_interface_2014, jin_tuning_2015, hastuti_theoretical_2024, liu_dynamic_2025}.
        This interface response is naturally included in first-principles calculations, but is absent from non-self-consistent tight-binding descriptions.
        We therefore introduce a minimal interface on-site correction on the top chalcogen layer closest to graphene and determine how this correction modifies the SOC parameters.

        We separately investigate the role of band alignment, for which different DFT studies and experiments report different values.
        The band alignment can also be modified by an applied electric field~\cite{zollner_twist-_2023, naimer_twist-angle_2021, yu_graphenemos2_2014, island_spinorbit-driven_2019, gerber_tunable_2025, dulisch_electric-field-tunable_2025, tiwari_electric-field-tunable_2021, gmitra_trivial_2016, jin_tuning_2015, hastuti_theoretical_2024, baik_work_2017}.
        Previous studies have reported a displacement of the graphene Dirac cone within the \tmd\ band gap of approximately $100$ meV per $V/nm$~\cite{zollner_twist-_2023, naimer_twist-angle_2021, rockinger_tuning_2026, szentpeteri_increasing_2025}.
        Here, we go beyond this tunable limit and investigate the full range of band alignments by shifting the graphene Dirac point throughout the \tmd\ band gap.

        We find that the local charge-transfer dipole correction produces only small changes in the SOC parameters.
        In contrast, the band alignment strongly affects the spin-valley and \rashba\ spin-orbit energies as well as the twist-angle dependence of the \rashbap.
        The response is material dependent, demonstrating that electrostatic alignment, in combination with the specific choice of \tmd, can account for part of the spread in the reported SOC values.
        
        The paper is organized as follows: in section~\ref{sec:tbmodel} we describe the TB model, after which we examine the effects of the charge-transfer dipole and band alignment on the parameters in sections~\ref{sec:dipole} and~\ref{sec:ba}, respectively, and compare the results with the literature.
        Finally, we provide a summary and outlook in section~\ref{sec:conclusions}.

    \section{Tight-Binding Model}\label{sec:tbmodel}
        \begin{figure}
            \centering
            \includegraphics[width=.35\linewidth]{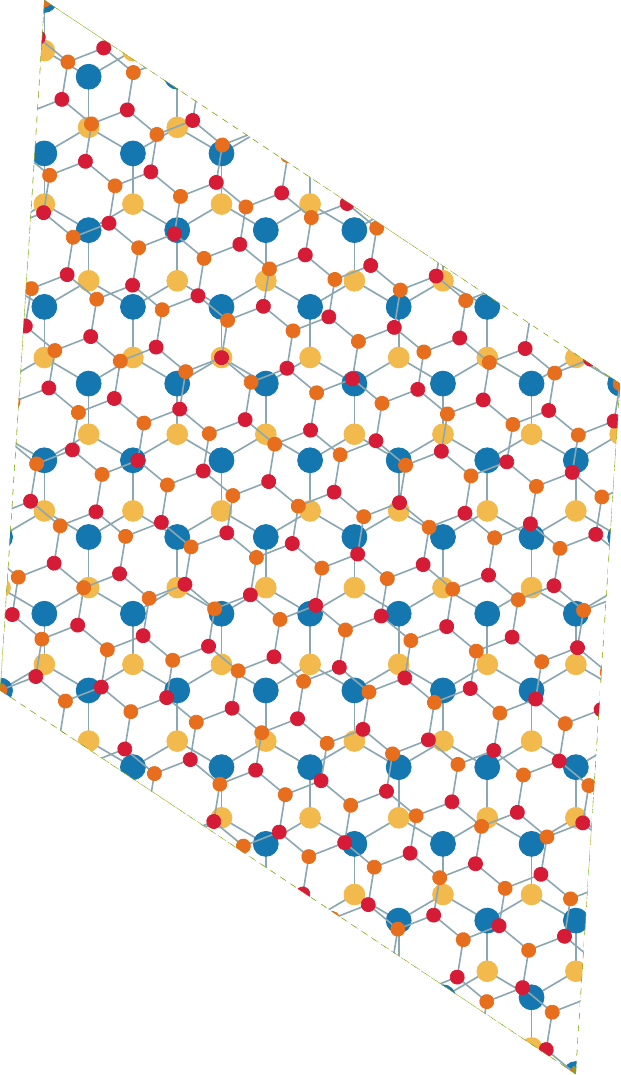}
            \caption{The supercell for a twisted graphene/\mos\ configuration with $\fo{7,-5,4,-5}$ and $\rottot=9.17\degre$, $\straintot=-3.28\%$. The orange and red sites indicate the graphene $\suba$ and $\subb$ sublattices, whereas the blue and yellow sites indicate the \tmd\ metal ($Mo$) and chalcogen ($S$) sites.}
            \label{fig:supercell}
        \end{figure}
        
        We construct the full tight-binding model for the heterostructure as a combination of a $2$-band monolayer graphene model $\hat H_{G}$~\cite{jung_accurate_2014} and an $11$-band monolayer \tmd\ model $\hat H_{\tmd}$~\cite{fang_ab_2015, fang_electronic_2018}, connected with an interlayer coupling term between the layers $\hat H_{G-\tmd}$~\cite{li_twist-angle_2019}.
        The Hamiltonian for the full TB model of the heterostructure is written as
        \begin{align}
            \hat H_{total} = \fo{\begin{array}{cc}
                \hat H_{\tmd} & \hat H_{G-\tmd} \\
                \hat H_{G-\tmd}^\dagger & \hat H_{G}
            \end{array}}.
        \end{align}

        The supercell for the full TB model is constructed from a specific combination of unit cells for the graphene and the \tmd, $\fo{\nG,\mg,\nt,\mt}$.
        The integer values $\left(\nG,\mg\right)$ determine the supercell vector for graphene, $\fo{\nt,\mt}$ for the \tmd.
        These values also determine the relative twist angle $\rottot$ and strain $\straintot$ between the layers, as elaborated in \apxref{sec:apx:construction_supercell}.
        For example, the configuration $\fo{7,-5,4,-5}$ for graphene on \mos\ results in the supercell shown in \figref{fig:supercell}, with twist angle $\rottot=9.17\degre$ and a strain of $\straintot=-3.28\%$.

        \subsection{Graphene and TMD Models}
        
        For graphene $\hat H_G$, we use a nearest neighbour hopping parameter $t=-2.61$ eV, lattice constant $\ag=0.246$ nm ~\cite{jung_accurate_2014}, on-site energy $\onsg=0$ eV and one $p_z$ orbital per site, leading to a 2-band model.
        The \tmd\ model $\hat H_{\tmd}$ is an 11-band model, including the $p$-orbitals on the two chalcogen atoms and the $d$-orbitals of the metal atom.
        The number of orbitals doubles with the inclusion of spin.
        In the \tmd\ TB model, the full SOC term $\mathbf{L}\cdot\mathbf{S}$ is included~\cite{jorissen_comparative_2024}.
        The \tmd\ TB model is transformed from a basis defined by the mirror plane along the metal atoms $\sigma_h$~\cite{fang_ab_2015,fang_electronic_2018} to atomic orbitals which are located at the sites of the two chalcogen atoms.

        \subsection{Interlayer Connection}
        \begin{figure}
            \centering
            \includegraphics[width=.57\linewidth]{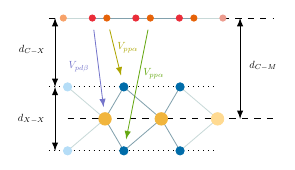}
            \includegraphics[width=.42\linewidth]{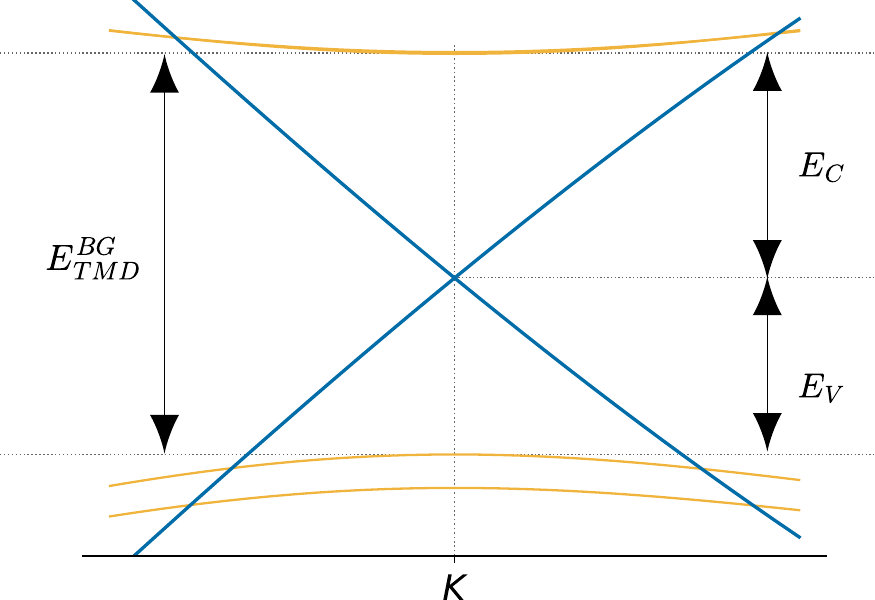}
            \caption{(left) The side view of the two layers in the heterostructure, with the relevant interlayer hoppings and distances indicated. (right) Schematic of the band alignment of the graphene Dirac cone in the \tmd\ band gap.}
            \label{fig:dirac_bg}
        \end{figure}
        
        The interlayer connection $\hat H_{G-\tmd}$ uses the Slater-Koster~\cite{slater_simplified_1954} description with an exponential decay term~\cite{li_twist-angle_2019, david_induced_2019, alsharari_mass_2016, peterfalvi_quantum_2022, wang_spinorbit_2021, singh_proximity-induced_2019} for the hopping from the graphene to the chalcogen ($V_{pp\alpha}\left(\mathbf{r}\right)$) and metal ($V_{pd\beta}\left(\mathbf{r}\right)$) atoms:
        \begin{align}\label{eq:hopping_strength}
            V_{pp\alpha}\fo{\mathbf{r}}&=V_{pp\alpha}^0e^{-\frac{|\mathbf{r}|-\dcx}{r^0_{\gc-\tsx}}},&
            V_{pd\beta} \fo{\mathbf{r}}&=V_{pd\beta}^0 e^{-\frac{|\mathbf{r}|-\dcm}{r^0_{\gc-\tsm}}},
        \end{align}
        where $\dcm$ and $\dcx$ are the vertical distances between graphene and the metal/chalcogen planes, respectively, and $\dxx$ is the vertical distance between the two chalcogen planes.
        The parameter $V_{pp\alpha}^0$ denotes the reference hopping amplitude between graphene $p$ and chalcogen $p$ orbitals, with $\alpha=\left\{\pi,\sigma\right\}$ the type of bond, and $V_{pd\beta}^0$ denotes the corresponding amplitude between the graphene $p$ and metal $d$ orbitals, with $\beta=\left\{\pi,\delta\right\}$.
        The parameters $r_{\gc-\tsm}^0$, $r_{\gc-\tsx}^0$ determine the respective decay lengths of the hopping amplitudes.
        The values of these parameters are given in \tabref{tab:interlayer_parameters}, and the hoppings are indicated in \figref{fig:dirac_bg}(a).

        \begin{table}
            \centering
            \begin{tabular}{rlr|llll}
                                  &        &                                 & \mos                                     & \mose                                    & \ws                                      & \wse                                      \\ \hline
                $\at$             & [$nm$] &\cite{fang_ab_2015}              &   0.318                                  &   0.332                                  &   0.318                                  &   0.332                                   \\
                $\dcx$            & [$nm$] &                                 &   0.337 \cite{gong_enhanced_2018}        &   0.341 \cite{ma_first-principles_2011}  &   0.341 \cite{kaloni_quantum_2014}       &   0.342 \cite{kaloni_quantum_2014}        \\
                $\dxx$            & [$nm$] &\cite{fang_ab_2015}              &   0.313                                  &   0.334                                  &   0.314                                  &   0.335                                   \\
                $\bacz$           & [eV] &\cite{gmitra_trivial_2016}       &    0.04                                  &    0.92                                  &    0.30                                  &    1.15                                   \\
                $\bavz$           & [eV] &                                 &    1.66                                  &    0.54                                  &    1.40                                  &    0.24                                   \\
                $\bgt$            & [eV] &                                 &    1.70                                  &    1.46                                  &    1.70                                  &    1.39                                   \\
                $\bafz$           & [$/$]  &                                 &    0.98                                  &    0.37                                  &    0.82                                  &    0.18                                   \\
                $r_{C-X}^0$       & [$nm$] &\cite{li_twist-angle_2019}       &  0.1131                                  &  0.1205                                  &  0.1131                                  &  0.1205                                   \\
                $r_{C-M}^0$       & [$nm$] &\cite{li_twist-angle_2019}       &  0.1280                                  &  0.1280                                  &  0.1367                                  &  0.1367                                   \\
                $V_{pp\sigma}^0$  & [eV] &\cite{li_twist-angle_2019}       &  1.6340                                  &  1.3271                                  &  1.5776                                  &  1.3161                                   \\
                $V_{pp\pi}^0$     & [eV] &\cite{li_twist-angle_2019}       & -0.4630                                  & -0.3766                                  & -0.4477                                  & -0.3735                                   \\
                $V_{pd\sigma}^0$  & [eV] &\cite{li_twist-angle_2019}       & -0.2674                                  & -0.2387                                  & -0.0866                                  & -0.0796                                   \\
                $V_{pd\pi}^0$     & [eV] &\cite{li_twist-angle_2019}       & -0.1544                                  & -0.1378                                  & -0.0500                                  & -0.0459                                   \\
            \end{tabular}
            \caption{Parameters used in the heterostructure calculations: the \tmd\ lattice constant $\at$, the interlayer distance $\dcx$, the vertical inter-chalcogen distance $\dxx$, the reference energies, $\bavz$ and $\bacz$, the \tmd\ band gap $\bgt$, the interlayer hopping decay lengths, $r_{C-X}^0$ and $r_{C-M}^0$, and hopping amplitudes $V_{pp\sigma}^0$, $V_{pp\pi}^0$, $V_{pd\sigma}^0$ and $V_{pd\pi}^0$.}
            \label{tab:interlayer_parameters}
        \end{table}

        \subsection{Full Model}
        The band alignment between graphene and \tmd\ ($\bac$) and the interlayer distance ($\dcx$) are important parameters of the full TB model. 
        The former specifies $\bac$, the energy difference between the graphene Dirac point and the conduction band minimum of the \tmd.
        Correspondingly, $\bav$ denotes the separation between the Dirac point and the valence band maximum,
        \begin{align}
            \bgt=\bac+\bav,
        \end{align}
        as illustrated in \figref{fig:dirac_bg}.
        The reference values for the band alignment obtained from the literature are indicated with a superscript, $\bacz$ and $\bavz$, whereas $\bac$ and $\bav$ denote the general values used when varying the position of the Dirac point within the \tmd\ band gap.

        To compare the band alignment across the different \tmd s, it is useful to introduce the fractional band alignment $\baf$:
        \begin{align}
            \baf &= \frac{\bav}{\bgt},
        \end{align}
        which gives the relative position of the graphene Dirac cone in the \tmd\ band gap.
        The interlayer distance $\dcx$ and the reference band alignment ($\bacz$, $\bavz$, $\bafz$) are given in \tabref{tab:interlayer_parameters}.
        
        \subsection{Effective Tight-Binding Model}
        To extract the proximity-induced SOC, we construct an effective tight-binding model $\hat H_G^{eff}$ that describes the influence of the \tmd\ on the graphene layer.
        This effective model has an additional term $\hat H_{G}^{SOC}$ which captures the proximity-induced SOC
        \begin{align}\label{eq:heffg}
            &\hat H_G^{eff} = -t\sum_{\langle i,j\rangle}\left(a_{i}^{\dagger}b_{j}+b_{j}^{\dagger}a_{i}\right) +  \ons \sum_i \left(\hat a_i^\dagger \hat a_i + \hat b_i^\dagger \hat b_i\right)  + \hat H_G^{SOC},
        \end{align}
        where $t$ is graphene's nearest-neighbour hopping, $\ons$ is the on-site energy and $\hat{a}\equiv(\hat{a}_{\uparrow},\hat{a}_{\downarrow})^{\text{t}}$ is the annihilation operator for the $A$ sublattice (similar definitions hold for sublattice $B$).
        The dominating piece of the SOC Hamiltonian, $\hat H_G^{SOC}$~\cite{perkins_spintronics_2024, kochan_model_2017, li_twist-angle_2019, gmitra_spin-orbit_2013, song_spin_2018, peterfalvi_quantum_2022, khatibi_proximity_2022, kang_magnetotransport_2024}, may be cast as
        \begin{align}\label{eq:heffsoc}
            \hat H_{G}^{SOC}&=\frac{i}{3\sqrt{3}}\lsv\sum_{\langle\langle i,j\rangle\rangle}\nu_{ij}\fob{\hat a^\dagger_i s_z \hat a_j - \hat b^\dagger_i s_z \hat b_j}\nonumber\\
                &\quad +\frac{2i}{3}\lr\sum_{\langle i,j\rangle}\fob{\hat a^\dagger_i e^{\frac{-i\prh s_z}{2}}\fo{\mathbf{s}\times\mathbf{d}_{ij}}_ze^{\frac{i\prh s_z}{2}}\hat b_j+\hat b^\dagger_i e^{\frac{-i\prh s_z}{2}}\fo{\mathbf{s}\times\mathbf{d}_{ij}}_ze^{\frac{i\prh s_z}{2}}\hat a_j},
        \end{align}
       where $\lsv$ and $\lr$ are the spin-valley and \rashba\ energies, respectively, $\prh$ is the \rashbap, and $\mathbf{d}_{ij}$ is the unit vector along the line segment connecting sites $i$ and $j$, with $\nu_{ij}=\pm1$ for anticlockwise ($+1$) and clockwise ($-1$) hopping.
        The effective model parameters are obtained by fits to data from the full TB model, as detailed in \apxref{sec:apx:fitting_routine}.

        The neglect of the Kane-Mele SOC term ($\lkm$) in our model is justified due to its smallness~\cite{lee_effective_2020, li_twist-angle_2019, david_induced_2019, naimer_twist-angle_2021, sichau_resonance_2019}.
        The (orbital) mass term, capturing the sublattice-staggered contribution to the on-site energy, $\Delta$, attaining values of the order of µeV, is also negligible (see \apxref{sec:apx:effective_model}) and shall not be considered further.
        For the \rashba\ SOC energy $\lr$, we only consider the absolute value $\lra$ as its sign is incorporated into the \rashbap\ $\prh$.
        In what follows, we also neglect pseudospin inversion asymmetry (PIA) SOC terms (described by the $\lpi$ and $\lpid$ spin-orbit energies~\cite{gmitra_spin-orbit_2013, gmitra_trivial_2016}) due to their small impact on energy band and spin expectation values for the parameter range of interest to us.
        The continuum description of the problem (including $\lkm$, $\Delta$, $\lpi$ and $\lpid$) is described in \apxref{sec:apx:effective_model}.

        \begin{table}
            \centering
            \begin{tabular}{cc}
                $0\degre/60\degre$                                                                & $30\degre$                                                              \\ \hline
                $\begin{array}{lll}
                \lsv\left(\rottot\right)&=\lsv\left(-\rottot\right)&=-\lsv\left(60\degre+\rottot\right)\\
                \lr\left(\rottot\right)&=\lr\left(-\rottot\right)&=\lr\left(60\degre+\rottot\right)\\
                \prh\left(0\degre\right)&=0\mod\pi&=\prh\left(60\degre\right)\mod 2\pi\\
                \ons\left(\rottot\right)&=\ons\left(-\rottot\right)&=\ons\left(60\degre+\rottot\right)\\
                \end{array}$&$\begin{array}{ll}
                \lsv\left(30\degre+\rottot\right)&=-\lsv\left(30\degre-\rottot\right)\\
                \lr\left(30\degre+\rottot\right)&=\lr\left(30\degre-\rottot\right)\\
                \prh\left(30\degre\right)&=0\mod\pi\\
                \ons\left(30\degre+\rottot\right)&=\ons\left(30\degre-\rottot\right)\\
                \end{array}$
            \end{tabular}
            \caption{Symmetries of the effective model.
            The parameters $\lr$, $\prh$ and $\ons$ have a period of $60\degre$, $\lsv$ of $120\degre$.}
            \label{tab:symmetrieseftb}
        \end{table}

        Spatial symmetries dictate the general behaviour of the effective spin-orbit energies in the effective model.
        The twisted heterobilayer has 3 vertical mirror planes for the special angle rotations of $30\degre$ and $60\degre$ (modulo $180\degre$).
        In such cases, the point symmetry group $C_{3v}$ of the untwisted system is respected.
        Moreover, for $\rottot=30\degre$, the spin-valley term $\lsv$ vanishes due to an emergent  sublattice symmetry~\cite{li_twist-angle_2019, perkins_spin_2024}.
        Note that for non-trivial twists, i.e. $\rottot \neq \{0\degre,30\degre$,60$\degre\}$ (module 180$\degre$), the mirror symmetries are broken and consequently the point group is lowered to $C_3$.
        A by-product of these considerations is that (i) the parameters $\lsv$, $\lr$ and $\ons$ exhibit mirror symmetry around $0\degre$; and (ii) around $30\degre$, $\lr$ and $\ons$ have another mirror symmetry, while $\lsv$ displays an inversion symmetry.
        Moreover, the phase of the \rashba\ parameter has to be $\prh=0 \mod \pi$ at the twist angles $\rottot=\{0\degre,30\degre,60\degre\}$ mod $180\degre$ ~\cite{li_twist-angle_2019, peterfalvi_quantum_2022, naimer_tuning_2024, naimer_twist-angle_2021, pachoud_scattering_2014}.
        An overview of the symmetries of these parameters is given in \tabref{tab:symmetrieseftb}. 

        \begin{figure*}
            \centering
            \includegraphics[width=\linewidth]{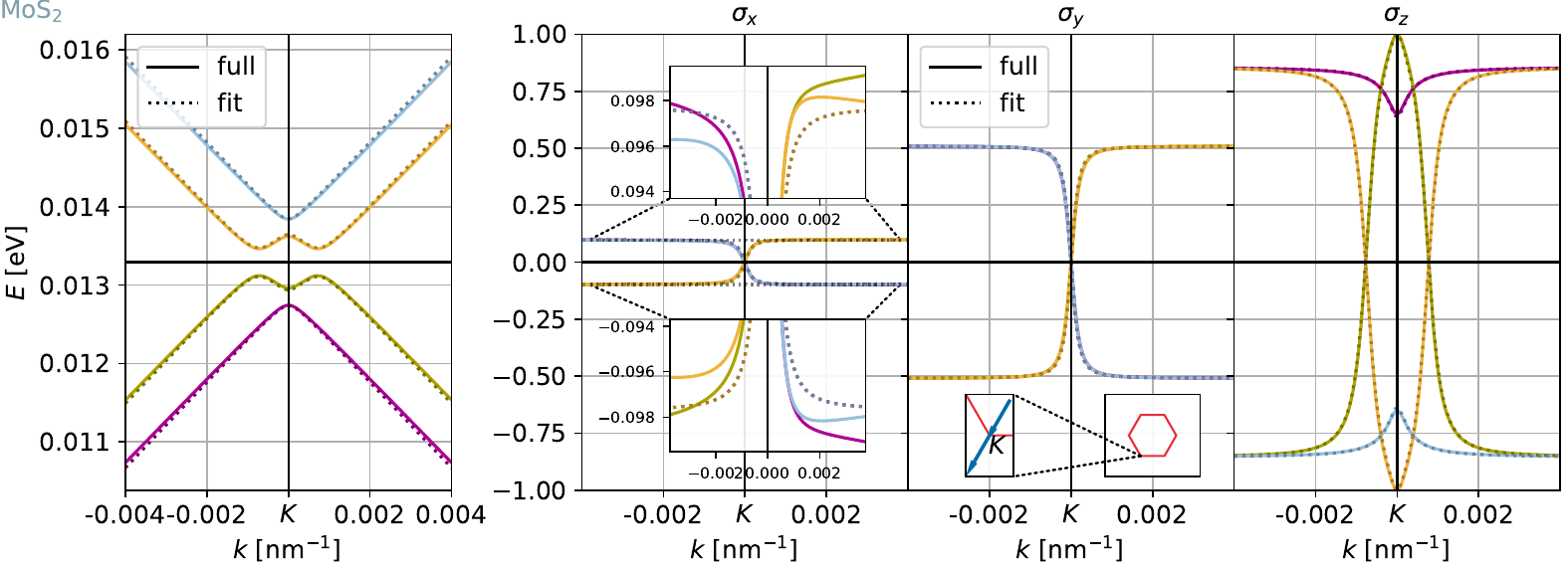}
            \caption{(left) Low-energy bands of the full (solid) and effective (dotted) TB model for \mos\ with $(7,-5,4,-5)$ as shown in \figref{fig:supercell}, with $\rottot=9.17\degre$, $\straintot=-3.28\%$ and $\bac=0.04$ eV.
            The fitted parameters are $\lsv=0.35$ meV, $\lr=0.21$ meV, $\prh=-1.24$ rad and $\ons=13.29$.
            (right) Spin expectation values for the low-energy bands.
            The $k$-path and first Brillouin zone are indicated in the inset to the middle panel.
            }
            \label{fig:low_energy_bands_spin}
        \end{figure*}
        
        In \figref{fig:low_energy_bands_spin}, the low-energy bands are shown for the full and fitted effective tight-binding models for the twisted graphene/\tmd\ heterostructure illustrated in \figref{fig:supercell}.
        One sees that the proximity-induced SOC has two pronounced effects: (i) a sizeable spin-splitting of the Dirac bands, and (ii) the avoided crossings of these spin-polarized bands (with associated spin-orbit gap), in accord with previous studies~\cite{offidani_optimal_2017,zollner_first-principles_2025,li_twist-angle_2019}.
        
        The on-site term $\ons$ represents the energy shift of the graphene Dirac cone in the effective model, induced by the proximity of the \tmd.
        The parameter $\prh$ (defined modulo $2\pi$) mainly influences the in-plane component of the momentum-space spin texture.
        The \rashbap\ $\prh$ in fact parameterises a $SU$(2) rotation about the $z$-axis, see \eqnref{eq:heffsoc}.
        When $\phi_R= 0$ the in-plane spin polarisation vector winds around the Fermi surface (spin-momentum locking), as expected from a Rashba SOC.
        Upon a $z$-rotation through some non-zero $\phi_R$, the Fermi-surface spin texture acquires a radial component.
        At the special Rashba angle $\phi_R=\pi/2$, the spin aligns with the electron momentum and the spin texture becomes purely radial \cite{perkins_spintronics_2024}.
        
        From \figref{fig:low_energy_bands_spin}, it is clear that the effective TB model provides a rather accurate representation of energy bands and momentum-space spin expectation values of graphene-TMD heterostructures.
        A detailed explanation of the fitting routines used in this work is given in \apxref{sec:apx:fitting_routine}.
        Armed with this understanding, in what follows we explore the rich twist-angle dependence of proximity-induced SOC as well as the role played by electrostatic tuning.
        
        \subsection{Twist-Angle Dependence}
        \begin{figure*}
            \centering
            \includegraphics[width=\linewidth]{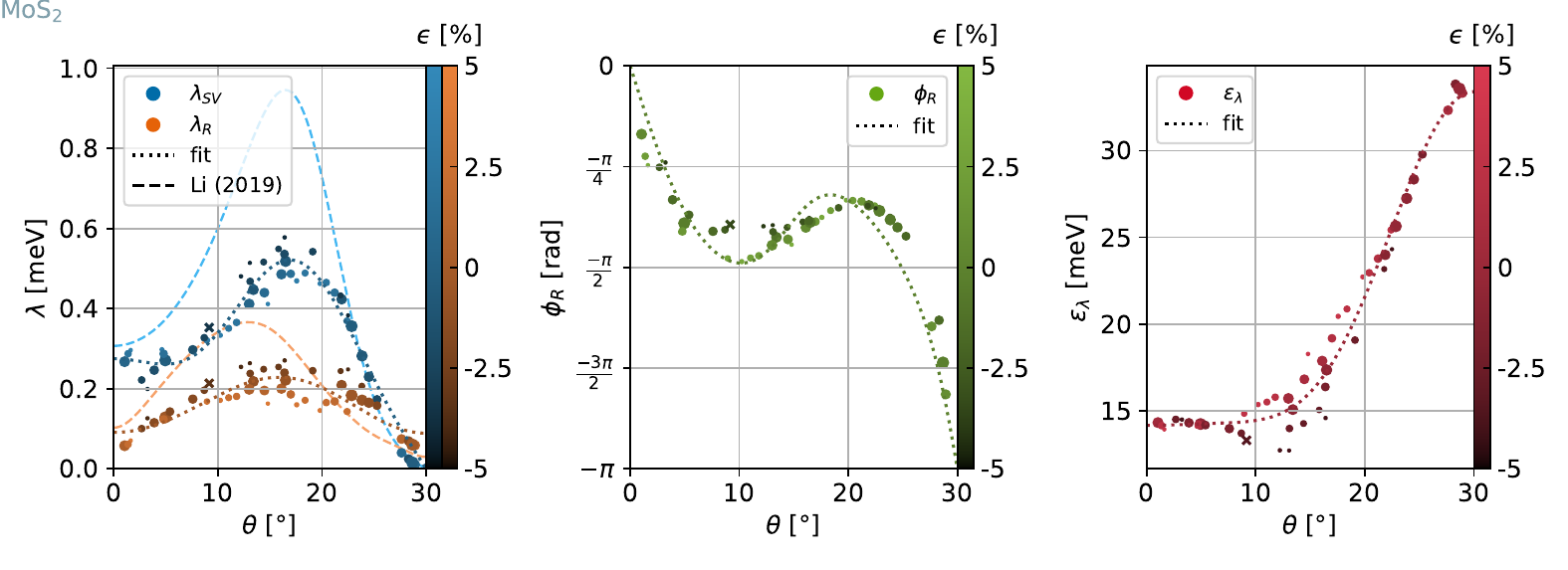}
            \caption{The SOC parameters for \mos, without dipole, with $\bacz=0.04$ eV offset.
            The cross indicates the configuration shown in \figref{fig:low_energy_bands_spin}.
            The harmonic fit is indicated with a dotted line.
            The colour and size of the dots correspond to the strain applied (larger dots indicate less strained structures).
            The results from \lk~\cite{li_twist-angle_2019} are indicated by the dashed lines on the left panel (note that they used a different band alignment $\baf$ and methodology).}
            \label{fig:parameters_angle}
        \end{figure*}

        We calculate the low-energy bands for different twist angles $\rottot$.
        Each twist angle corresponds to a specific configuration $\left(\nG,\mg,\nt,\mt\right)$.
        We select configurations with a small number of orbitals, $N_{orbitals}$ (less than 2000).
        Similar configurations are then filtered by requiring a positive twist angle $\rottot$ and a small graphene rotation $|\rotlayg|$ (see \apxref{sec:apx:construction_supercell}).
        The systems are selected such that the total strain in the graphene layer is $|\straintot| <5\%$.
        
        The resulting parameters for graphene on \mos\ with different $\left(\nG,\mg,\nt,\mt\right)$ configurations, and thus the twist angle $\rottot$, are shown in \figref{fig:parameters_angle}.
        Each configuration is indicated by a dot, where the dot size and colour represents the strain applied to the graphene layer.
        
        We perform a harmonic fit to obtain a twist-angle dependence of the effective SOC parameters.
        In this harmonic fit, we determine the coefficients for specific $\sin$-functions that share the periodicity of the target spin-orbit energy, as explained in \apxref{sec:apx:fitting_routine}.
        The fit covers quite well the different characteristics of the parameters with the variation of the twist angle.
        The \rashbap\ $\prh$ varies from $0$ to $-\pi$, with a small region where $\prh=\frac{-\pi}{2}$ radians is observed around $\rottot=10\degre$.
        This enables a purely collinear Rashba-Edelstein effect as predicted in \refcite{veneri_twist_2022}.
        
        The spread of the calculated points around the fitted line in \figref{fig:parameters_angle} originates primarily from the strain in the graphene unit cell.
        The variation of strain will change the local environment experienced by low-energy carriers.
        The relative importance of the different interlayer hopping processes therefore changes with strain.
        We take variations in strain up to $\strainmax=\pm5\%$, which results in a variation for the unit cell area of $\pm10\%$.

     \section{Charge-Transfer Dipole}\label{sec:dipole}
        \begin{figure*}
            \centering
            \includegraphics[width=\linewidth]{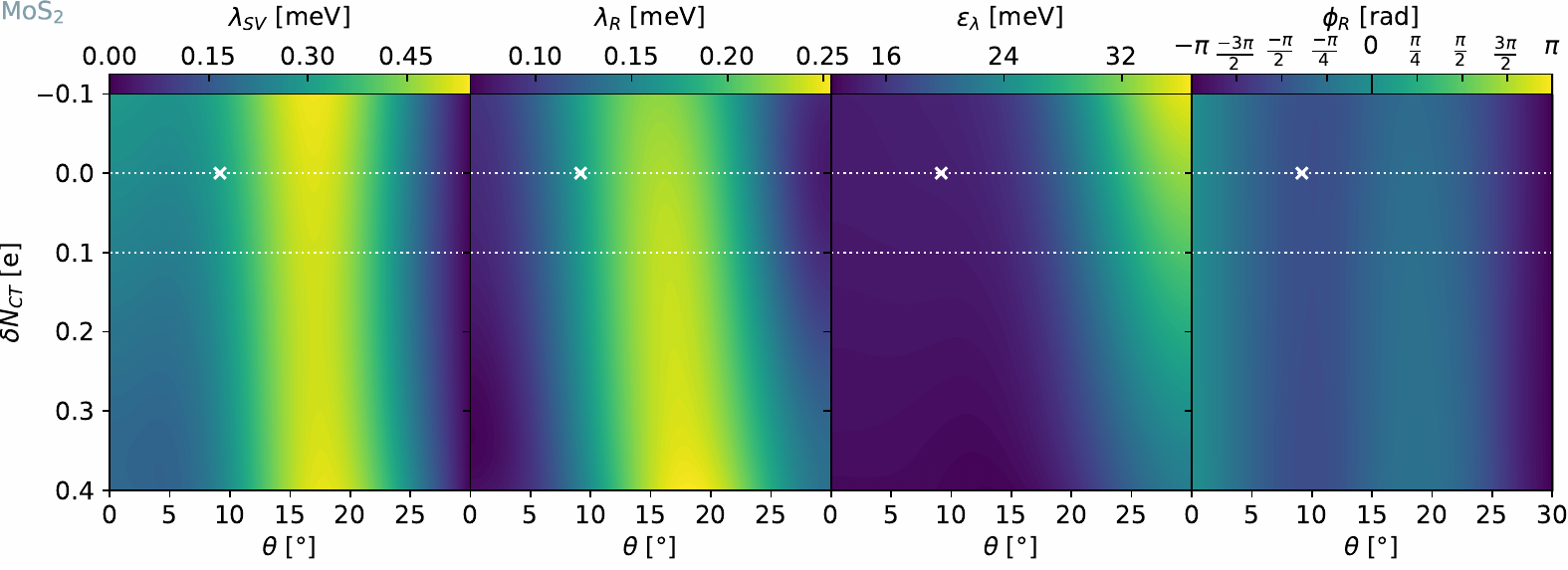}
            \caption{Changes to the SOC parameters with the tuning of the dipole, with the reference $\bac$ as given in \tabref{tab:interlayer_parameters} for \mos.
            Changing the dipole does not produce significant changes in the parameters.
            The configuration from \figref{fig:low_energy_bands_spin} and the dipole of $\dct= 0 e$ and $+0.1 e$ are indicated with a cross and dotted lines, respectively.
            }
            \label{fig:dipole}
        \end{figure*}
        An interface dipole shifts the on-site energies of the \tmd\ and graphene in opposite directions.
        Because a shift of the graphene on-site energy is equivalent to a change in band alignment, we consider only the contributions from the \tmd\ for the charge-transfer dipole.
        The charge-transfer dipole on the \tmd\ is investigated using a minimal model with a change in the on-site energy for the top chalcogen layer:
        \begin{equation}
            \oxt = \oxtz + \oxtct\fo{\dct}
        \end{equation}
        with $\oxtz$ the on-site energies for the top chalcogen atoms closest to the graphene layer.
        The change in on-site energy $\oxtct\fo{\dct}$ is correlated to the additional filling $\dct$ of the orbitals in the chalcogen orbitals closest to the graphene, as explained in \apxref{sec:apx:dipole}.
        In the literature, various values for this charge transfer from graphene to \mos\ have been reported, ranging from negligible~\cite{pierucci_band_2016, thi-xuandang_interface_2025} to maximal $\dct\approx 0.1 e$ (electrons per unit cell) ~\cite{sukhanova_induced_2020, yun_electronic_2025, fang_van_2022}, although higher values have been reported for doped systems~\cite{thi-xuandang_interface_2025} and specific substrates~\cite{dappe_charge_2020}.

        The twist-angle dependence and the charge-transfer-dipole dependence of the parameters are investigated simultaneously.
        We use a two-step approach to obtain this dependence on two variables: first, we obtain harmonic fits similar to those in \figref{fig:parameters_angle} for different values of the charge-transfer dipole, after which the harmonic parameters from the first fitting routine are themselves fitted as a function of the charge-transfer dipole with a polynomial fit.
        The resulting variation of the effective model parameters with both a change in twist angle and charge-transfer dipole for graphene on \mos\ is shown in \figref{fig:dipole}.
        This fitting routine is explained in detail in \apxref{sec:apx:fitting_routine}.
        As a baseline for the charge-transfer dipole filling, we take $\dct=+0.1 e$ for all heterostructures.
        
        The charge-transfer dipole correction does not produce a significant change in the SOC, an increase of approximately $5\%$ is observed for $\lr$ between $\dct=0e$ and $\dct=.1 e$.
        Additionally, the on-site parameter $\ons$ varies, which is expected because the dipole modifies the on-site energies of the full model and therefore affects the fitted on-site parameter of the effective model.
        Similar results obtained for \ws, \mose\ and \wse\ are shown in appendix \apxref{sec:apx:more_results}, with slightly larger variations in the parameters observed in \mose\ and \wse.
        
    \section{Change of Band Alignment}\label{sec:ba}
        \begin{figure*}
            \centering
            \includegraphics[width=\linewidth]{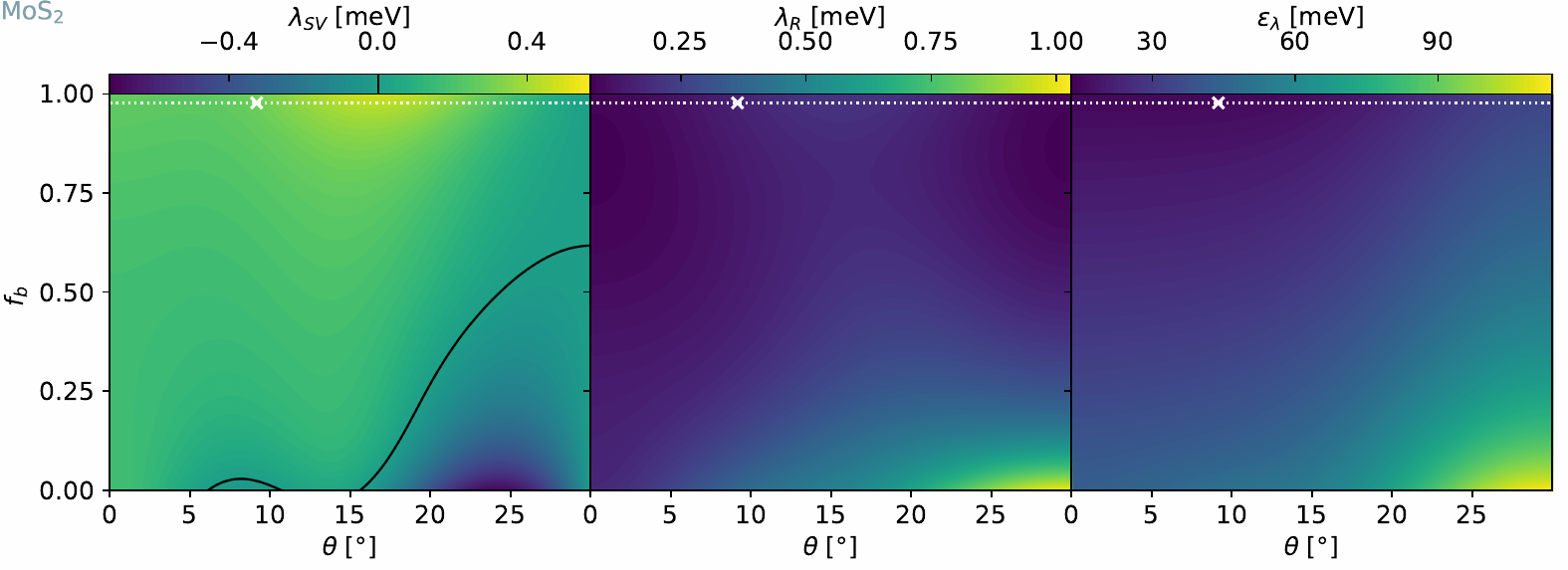}
            \caption{The SOC parameters with a change in band alignment for \mos. The harmonic fit from \figref{fig:parameters_angle} and calculation from \figref{fig:low_energy_bands_spin} are indicated with a dashed line and cross respectively. No charge-transfer dipole is applied in the calculation, $\oxtct=0eV$.
            A black line indicates the region where the parameters are $0 eV$.}
            \label{fig:ba_params_mos}
        \end{figure*}
        
        Changes in band alignment are incorporated through a shift in the on-site energies of graphene
        \begin{equation}
            \onsg = \onsgz + \onsgs
        \end{equation}
        with $\onsgz$ being the reference on-site energy of graphene without the change in band alignment and $\onsgs$ the additional shift on the graphene on-site energies, such that the on-site energy in the full TB calculation is $\onsg$.
        Without the change in band alignment, the on-site energy of graphene is set such that the Dirac cone has the band alignment as mentioned in \tabref{tab:interlayer_parameters}, $\onsgz\rightarrow\bavz$.
        Tuning the band alignment is equivalent to shifting the Dirac point within the \tmd\ band gap, $\onsg\rightarrow\bav$.
        To better compare the SOC parameters across materials, we use the relative shift in the band gap $\baf$.
        In \figref{fig:ba_params_mos}, the parameters for \mos\ are given with a variation of the band alignment.
    
        The spin-valley parameter $\lsv$ decreases in value and becomes negative as the Dirac point approaches the valence band, while the \rashba\ parameter $\lr$ and on-site term $\ons$ increase in value closer to the valence bands.
        It is clear that the band alignment has an influence on the low-energy parameters, with both the value and the location of their extrema varying.
        Similar results for \mose, \ws\ and \wse\ are shown in the \apxref{sec:apx:more_results}, where also the results for simultaneous variations in band alignment and dipole are provided.
        
        \begin{figure*}
            \centering
            \includegraphics[width=\linewidth]{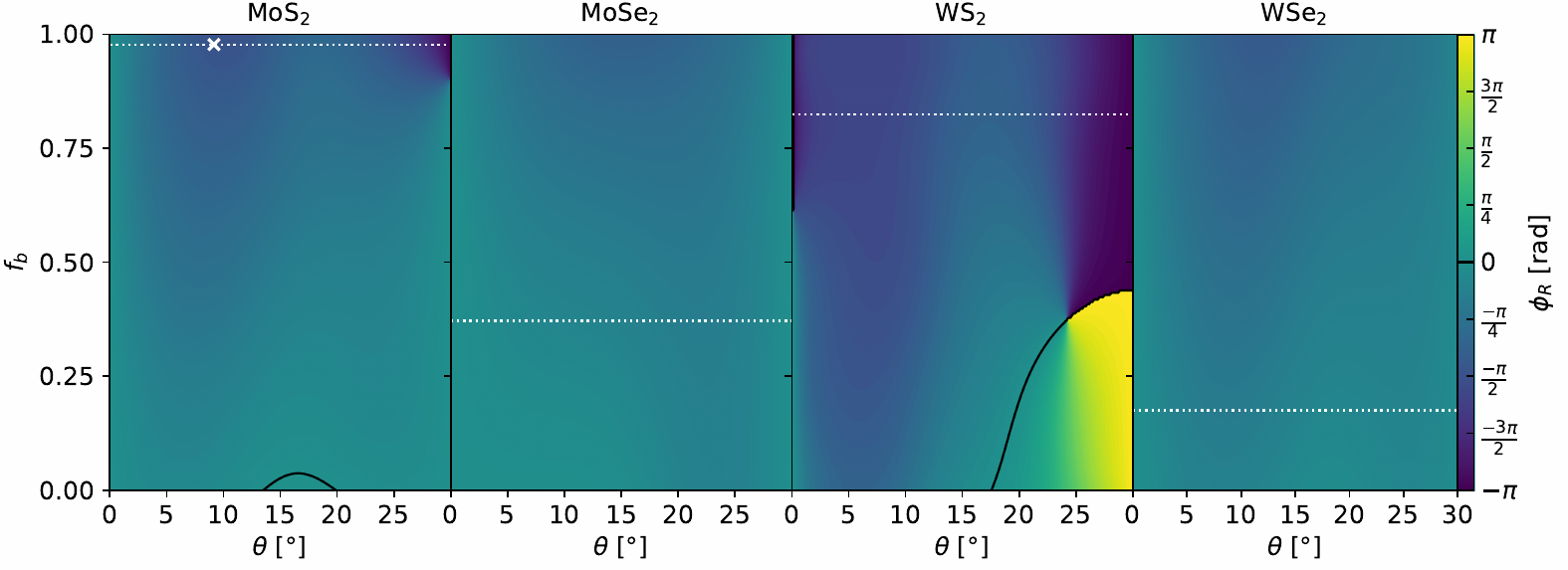}
            \caption{The \rashbap\ for the TMD materials, with the dashed line indicating the reference $f_{b}$ given in \tabref{tab:interlayer_parameters}. No dipole is applied in the calculation.
            The set of points where $\prh$ is $0$ or $\pm\pi$ are indicated with a black line.}
            \label{fig:ba_phi}
        \end{figure*}
        
        The \rashbap\ $\prh$ across the four \tmd s is compared in \figref{fig:ba_phi}.
        For \mos, there is a switch around $\baf=0.93$ at $\rottot=30\degre$, where the \rashbap\ goes from $\prh=-\pi$ above $\baf=0.93$ to $\prh=0$ below this filling.
        However, for \mose\ and \wse, no such behaviour is observed, in both cases the \rashbap\ goes from $\prh=0$ at $\rottot=0\degre$ to $\prh=0$ at $\rottot=30\degre$.
        For \ws, the \rashbap\ switches two times.
        First, a switch at around $\baf=0.66$, where at $0\degre$ twist angle $\prh=-\pi$ goes to $\prh=0$ above this filling.
        Then, there is a second switch around $\baf = 0.45$, where at $30\degre$ twist angle $\prh=-\pi$ changes to $\prh=\pi$.
        More details about this switching behaviour and the fit of $\prh$ are given in \apxref{sec:apx:phir_ws}.
                
        \begin{figure}
            \centering
            \includegraphics[width=\linewidth]{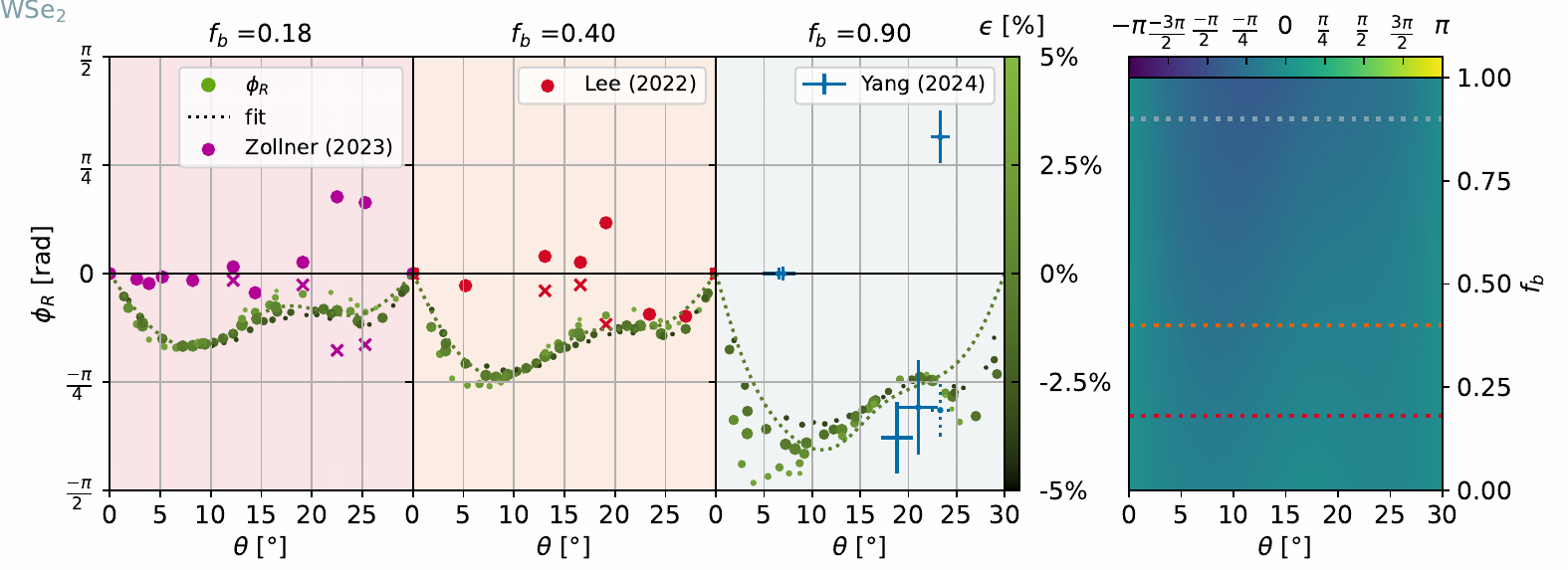}
            \caption{Different values of $\prh$ predicted for \wse. The points indicated by a cross and dotted error bars represent the negative of the corresponding reported values.}
            \label{fig:phircomp}
        \end{figure}
        
        Our results show that the \rashbap\ exhibits different behaviour depending on the specific band alignment $\baf$.
        By symmetry, $\prh$ can only take values that are integer multiples of $\pi$ at the twist angles of $0\degre$ or $30\degre$.
        This switching behaviour for $\prh$ was observed before~\cite{peterfalvi_quantum_2022}.
        \textit{P\'eterfalvi et al.} highlight that the change in $\prh$ they observed is related to the band alignment.
        Comparison with theoretical and experimental results from literature gives noticeable quantitative differences, as shown in \figref{fig:phircomp}.
        For twist angles between approximately $20\degre$ and $30\degre$, some agreement is observed with the calculations of \textit{Zollner et al.} and \textit{Lee et al}.
        The extraction of $\prh$ is not straightforward, as this is an effect captured by the spin expectation values, which complicates accurate calculations.
        Moreover, obtaining reliable measurements for $\prh$ in experiments is not straightforward either but a possible approach relies upon disentangling standard (tangential) and collinear Rashba-Edelstein responses by means of lateral spin-valve measurements in a tilted magnetic field~\cite{cavill_proposal_2020}.
        Encouraging experimental results for graphene/\wse\ heterostructures have been reported recently by \textit{Yang et al.}~\cite{yang_twist-angle-tunable_2024}.
        The rich diagram for \ws\ is also compared with literature, as elaborated in \apxref{sec:apx:phir_ws}.
        We conclude that the changes for the \rashbap\ are mainly due to a change in band alignment.
        
        The band alignment can be tuned by applying an electric field.
        Previous studies investigated the change in electric field~\cite{zollner_twist-_2023, naimer_twist-angle_2021, yu_graphenemos2_2014} and the influence of the position of the Dirac cone in the band gap of the TMD, with a tunability of $\sim100$ meV per $V/nm$~\cite{zollner_twist-_2023}.
        The experimentally accessible region for the electric field is of $E_{max}\sim\pm 1 V/nm$ ~\cite{island_spinorbit-driven_2019, gerber_tunable_2025, dulisch_electric-field-tunable_2025, tiwari_electric-field-tunable_2021}.
        In literature, it has been reported that a change in electric field~\cite{gmitra_trivial_2016, naimer_twist-angle_2021} leads to a large change in the SOC parameters.
        Combined with precise twist-angle control, specific materials can be created in which the parameters could be changed on-demand by tuning  the electric field.
        This hints that this \rashbap\ might be tunable, with a phase diagram similar to the one reported in \figref{fig:ba_phi}.
        Furthermore, the SOC parameters depend upon the interlayer distance and thus are tunable by pressure \cite{naimer_twist-angle_2021, zollner_twist-_2023, rockinger_tuning_2026, szentpeteri_increasing_2025}.

        The material specific results seen in \figref{fig:phircomp} and \apxref{sec:apx:ba_dipole} seem to give two trends, one depends on the metal ($Mo$/$W$) and the other one on the chalcogen ($S$/$Se$).
        The rich nature of the diagrams for $\prh$ seen in \mos\ and \mose\ indicate that there is a dependence of this behaviour on the chalcogen species.
        On the contrary, the results obtained for $\lsv$ and $\lr$ show a dependence on the metal, where the heavier metal $W$ leads to an overall stronger effect on the induced SOC parameters.
        
    \section{Conclusion}\label{sec:conclusions}
        We investigated how the electronic environment of the \tmd/graphene heterostructure influences the proximity-induced SOC in graphene on \mos, \mose, \ws\ and \wse.
        We focused on two main effects: a charge-transfer dipole and a change in the relative band alignment.
        We determined the twist-angle dependence of the SOC parameters by fitting an effective graphene Hamiltonian to the full tight-binding model.

        We used a minimal model to describe the charge-transfer dipole by shifting the on-site energies of the chalcogen atoms closest to the graphene layer.
        By separating the effects of the charge-transfer dipole on the \tmd\ and graphene layers, we distinguish the contributions of charge redistribution from a shift of the graphene Dirac cone, which is described through a change in band alignment.
        Over the investigated range for the charge-transfer dipole, only modest changes in the SOC parameters were observed.
        The \rashba\ parameter for graphene/\mos\ changes by approximately $5\%$ between $\dct=0e$ and $0.1e$.
        Therefore, the \tmd-side contribution of the dipole does not significantly change the calculated SOC parameters.
        
        In contrast, the position of the graphene Dirac point within the \tmd\ band gap has a significant influence on the SOC parameters.
        Varying the fractional band alignment $\baf$ changes both the magnitude and twist-angle dependence of the spin-valley and \rashba\ parameters.
        The band alignment also changes the character of the \rashbap.
        Depending on the material and band alignment, $\prh$ can switch between values of $0$ and $\pm\pi$ at twist angles of $0\degre$ and $30\degre$.
        The strong sensitivity of $\prh$ to $\baf$ provides a possible explanation for part of the variation among reported values.
        Because different samples and computational approaches can correspond to different band alignments, their electrostatic conditions must be considered when comparing results.

        The strong dependence of $\prh$ on the band alignment indicates the possibility of using electrostatic control to tune the in-plane spin texture of the heterostructure.
        The calculated twist-angle dependence for $\prh$ shows trends similar to the experimentally observed angular dependence for \wse, although a quantitative comparison remains sensitive to the parameter extraction method and to structural and electrostatic details that differ between samples.

        Several effects remain outside the scope of the presented model.
        The commensurate supercells require strain in the graphene layer, which modifies the interlayer hopping environment.
        However, the strain dependence of the graphene intralayer tight-binding parameters is not included.
        Atomic relaxation and moir\'e reconstruction can further modify the local stacking and interlayer distances~\cite{li_relaxation_2024, beule_lattice_2026, zhai_twistronics_2025}. 
        The parameterization obtained here nevertheless provides a basis for constructing spatially varying effective tight-binding models in which local twist angle, band alignment, and interface environment are incorporated through position-dependent parameters.
        
    \textit{Acknowledgements}
        The computational resources and services used in this work were provided by the HPC core facility CalcUA of the Universiteit Antwerpen, and VSC (Flemish Supercomputer Center), funded by the Research Foundation - Flanders (FWO) and the Flemish Government.
        We thank Robin Smeyers for the help in coding the interlayer hopping in our tight-binding model, and Ivan Verstraeten for the useful discussions about the graphene on TMD heterostructures.
    \paragraph{Funding information}
        Bert Jorissen acknowledges the support of the Flemish Research Foundation (FWO/11E5821N).

    \begin{appendices}
    \numberwithin{equation}{section}
    \section{Construction of the Model}\label{sec:apx:construction_supercell}
        The heterobilayer model is constructed by first obtaining accurate models for the monolayers, cutting out a specific area of this monolayer, rotating and straining the layers and finally adding interlayer hoppings.
        \subsection{Graphene Model}
            The unit cell for graphene is given by the two lattice vectors
            \begin{align}\label{eq:grapheneunit cell}
                \aone &=\ag\ux & \atwo &=\ag\left(-\frac{1}{2}\ux + \frac{\sqrt{3}}{2} \uy\right),
            \end{align}
            with $\ag$ the lattice spacing.
            The distance between the nearest-neighbour carbon atoms is given by $\acc=\frac{\ag}{\sqrt{3}}$.
            For monolayer graphene, the two sublattices are labelled $\suba$ and $\subb$.
            The positions of these sublattices are given by
            \begin{align}\label{eq:graphenePositions}
                \posvector_{\suba} &= \left(\begin{matrix}0\ux\\0\uy \end{matrix}\right),&
                \posvector_{\subb} &= \left(\begin{matrix}\frac{\ag}{2}\ux\\\frac{\ag\sqrt{3}}{6}\uy\end{matrix}\right).
            \end{align}
            
            At each of these positions, we consider one $p_z$-orbital.
            This orbital is split into a spin-up ($\uparrow$) and spin-down ($\downarrow$) component, doubling the number of   orbitals per unit cell from 2 to 4.
            Since the intrinsic SOC is neglected (see main text), the TB Hamiltonian for graphene is block diagonal in spin space.
            The nearest-neighbour hopping $t$ connects orbitals within the graphene layer with the same spin.
            We used the parameters reported by \textit{Jung \& MacDonald}~\cite{jung_accurate_2014}.
            
        \subsection{\tmd\ Layers}
            The unit cell for the \tmd s is similar to the one for graphene in equation~\eqref{eq:grapheneunit cell}, but with the unit cell parameter $\at$.
            In the \tmd, the metal ($\tsm$) and chalcogenides ($\tsx$) are located at $\suba$ and $\subb$ positions analogous to equation~\eqref{eq:graphenePositions}.
            In this work, we consider the following TMDs: \mos, \mose, \ws\ and \wse.
            The entire \tmd\ layer is shifted down in the $z$-direction with a distance $\dcm$.
            The top ($\tsxu$) and bottom $(\tsxl$) chalcogenides on the $\subb$ position are vertically separated by a distance $\dxx$.
            The positions of the sublattices are thus given by
            \begin{align}\label{eq:tmdPositions}
                \posvector_{\tsm} &=
                    \left(\begin{matrix}
                        0 \ux\\ 0 \uy\\ -\dcm\uz
                    \end{matrix}\right),\\
                \posvector_{\tsxu} &=
                    \left(\begin{matrix}
                        \frac{\at}{2}\ux \\ \frac{\at\sqrt{3}}{6}\uy \\ \left(-\dcm+\frac{\dxx}{2}\right)\uz
                    \end{matrix}\right), &
                \posvector_{\tsxl} &=
                    \left(\begin{matrix}
                        \frac{\at}{2}\ux \\ \frac{\at\sqrt{3}}{6}\uy \\ \left(-\dcm-\frac{\dxx}{2}\right)\uz
                    \end{matrix}\right).\nonumber
            \end{align}
            
            For the tight-binding description of the \tmd\, we used the parameters from \textit{Fang}~\cite{fang_ab_2015, fang_electronic_2018} implemented in Pybinding~\cite{moldovan_pybinding_2020, moldovan_bertjorissenpybinding_2026, harris_array_2020} within the TMDybinding expansion~\cite{jorissen_comparative_2024}.
            This amounts to a 11-band TB model, which includes the $p$-orbitals on the chalcogenides and the $d$-orbitals on the metal atoms.
            This basis is doubled to include the spin-up and -down components.
            The full $\mathbf{L}\cdot\mathbf{S}$-term for the intra-atomic SOC interaction in the \tmd\ is taken into account, which connects the spin-up and -down sectors.
            The basis orbitals are located at the atomic positions as given in equation \eqref{eq:tmdPositions}.
            Further details on the construction of the Hamiltonian for the \tmd, parameters and the implementation of the SOC terms can be found in~\cite{jorissen_comparative_2024}.

        \subsection{Connecting Graphene and \tmd}
            The interaction between graphene and \tmd\ is modelled by means of interlayer hoppings that are exponentially decaying with the distance~\cite{li_twist-angle_2019, david_induced_2019, alsharari_mass_2016, peterfalvi_quantum_2022, wang_spinorbit_2021, singh_proximity-induced_2019}, following a standard Slater-Koster approach~\cite{slater_simplified_1954, fang_ab_2015, fang_electronic_2016, venkateswarlu_electronic_2020, vitale_flat_2021}.
           
            The Slater-Koster hoppings read as
            \begin{align}\label{eq:apx:hop_amp}
                V_{pp\alpha}\left(\textbf{r}\right)&=V_{pp\alpha}^0e^{-\frac{|\mathbf{r}|-d_{C-X}}{r^0_{C-X}}},&
                V_{pd\alpha}\left(\textbf{r}\right)&=V_{pd\alpha}^0e^{-\frac{|\mathbf{r}|-d_{C-M}}{r^0_{C-M}}},
            \end{align}
           where $V^0_{pp\pi},V^0_{pd\sigma},V^0_{pd\pi}$ are constant prefactors, $d_{C-X/M}$ is reference bond length for a carbon-metal/chalcogen bond and $r^0_{C-X/M}$ is the respective decay length.
            In \lk\, an additional fitting constant $\eta$ is used, which we absorbed in the pre-factors.
            Furthermore, there is a dimensional inconsistency with their equation (5), where a scaling factor of $d^4$ should appear on the denominator of the last two equations in line with the Harrison rule~\cite{harrison_elementary_1999}.

            Other works have employed different approaches.
            For example, \refcite{david_induced_2019, peterfalvi_quantum_2022} obtain order-of-magnitude estimates of the hopping amplitudes between the $p_z$-graphene and the \tmd\ orbitals by comparison with the well-studied interlayer hopping parameter in bilayer graphene as the interlayer distances are similar in both cases.
            Another study uses the \tmd s intralayer Slater-Koster parameters for the interlayer hopping parameters of the heterostructure~\cite{singh_proximity-induced_2019}.
            Because these Slater-Koster parameters depend strongly on the orbital basis and parameterization of the underlying tight-binding model~\cite{jorissen_comparative_2024}, their numerical values cannot be transferred directly between different models.
            Nevertheless, these estimates provide a reasonable basis for qualitative calculations of the general behaviour of the systems.
            
            Our heterostructure parameters are given in \tabref{tab:interlayer_parameters}.
            A cutoff distance for the interlayer hoppings of 4 times the interlayer distance is employed.
            The constants $V_{pp\alpha}^0$ and $V^0_{pd\beta}$ have been rescaled from \cite{li_twist-angle_2019} to be in line with \eqnref{eq:hopping_strength}.
            The other parameters satisfy the relations \cite{li_twist-angle_2019} 
            \begin{align}
                 \frac{r^0_{C-X}}{\tilde d_{C-X}}&= 1.609 = \frac{r^0_{C-M}}{\tilde d_{C-M}},
            \end{align}
            with $r^0_{C-X}$ and $r^0_{C-M}$ the decay lengths as given in \tabref{tab:interlayer_parameters}, $\tilde d_{C-X}$ and $\tilde d_{C-M}$ are the reference bond lengths used in their paper, given by $\tilde d_{C-S}=0.182nm$~\cite{huheey_strengths_1958}, $\tilde d_{C-Se}=0.194nm$~\cite{jenks_photochemistry_1994}, $\tilde d_{C-Mo}=0.2059nm$~\cite{mak_crystal_1984} and $\tilde d_{C-W}=0.22nm$~\cite{rudy_untersuchungen_1962}.

            A quantitative study as a function of the interlayer distance is outside the scope of this paper.
            However, a general conclusion that emerges is that the low-energy electronic structure and spin textures are reliably captured by the effective graphene-only TB model.
            This conclusion is, of course, robust with respect to variations of interlayer hoppings.
            A possible venue for future work is to generate a parameterization of a more general and accurate TB model that can include local changes to the twist angle, strain or interlayer effects by modulating the TB parameters in real space. Such a study would complement \refcite{khatibi_proximity_2022}, where the impact of local defects in the \tmd\ was investigated.

        \subsection{Cutting the Layers}
            \begin{figure}
                \centering
                \includegraphics[width=\linewidth]{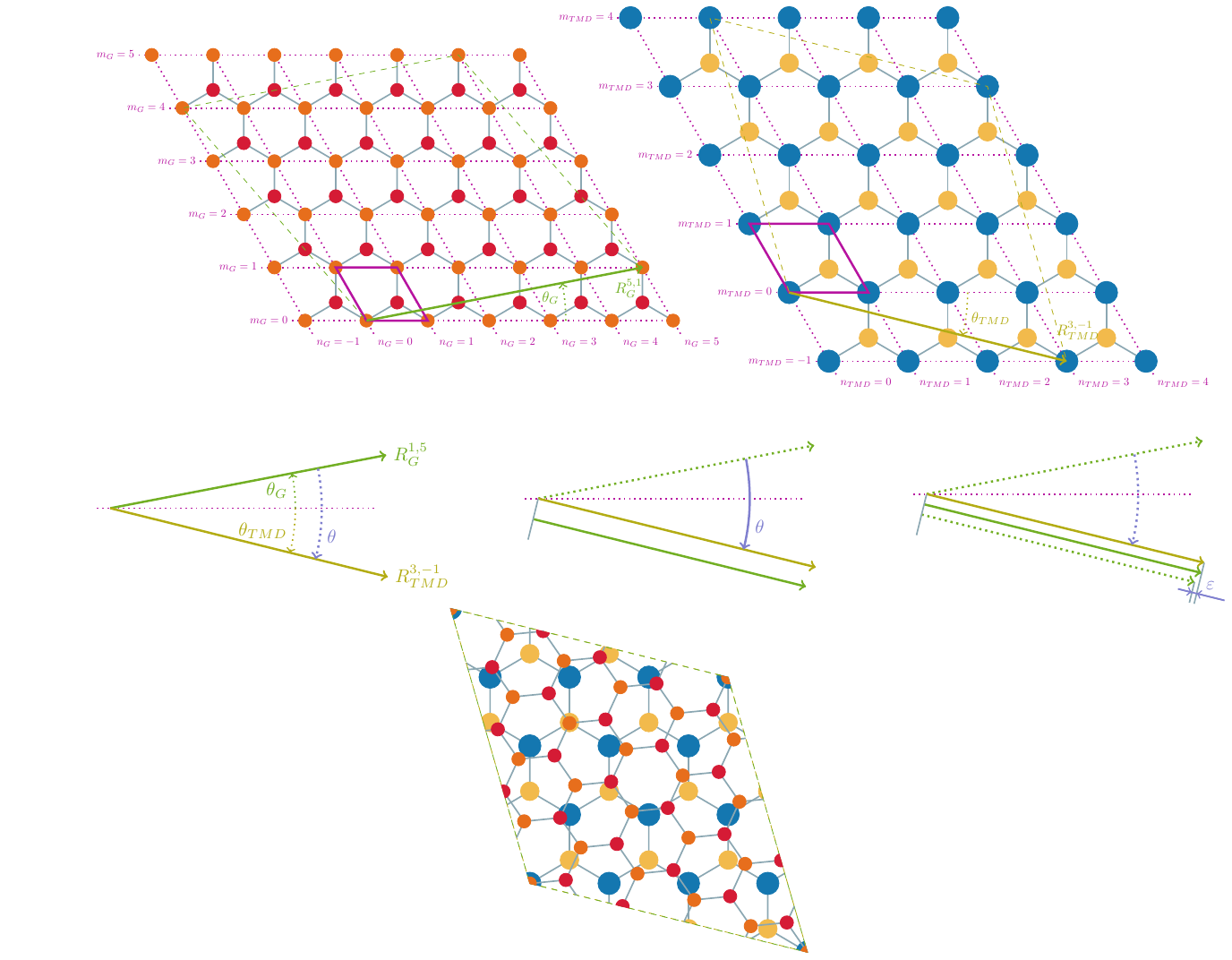}
                \caption{The construction of the commensurate supercell. (top) the vectors $\Rnm$ are constructed from the unit cells of the different monolayers, (middle left) the relative twist angle between the layers $\rottot$ is determined and (middle right) the relative strain $\straintot$ is calculated with the length difference of the two vectors, (bottom) the layers are rotated by $\rottot$ and strained with $\straintot$ to obtain the commensurate supercell.
                The example is taken for graphene on \mos\ with $(5,1,3,-1)$.}
                \label{fig:supercell_construction}
            \end{figure}
            Our tight-binding calculations require a commensurate supercell~\cite{carr_electronic-structure_2020, necio_supercell-core_2020, latil_structural_2023, gratias_crystallography_2023, koda_coincidence_2016, zollner_first-principles_2025}.
            At least one of the two layers will need to be strained to construct this supercell.
            In our calculation, the graphene layer is rotated and strained relative to the \tmd\ layer.
            
            The supercell for the heterostructure is created by first selecting a supercell in both layers.
            Each supercell is defined by the supercell cutting vector, $\Rgnm$ and $\Rtnm$.
            The second cutting vector obtained by rotating the first vector by $120\degre$, resulting in a cutout which is congruent to the original unit cell given by \eqnref{eq:grapheneunit cell}.
            The two supercell vectors $\Rnm$ are multiples of the unit cell vectors $\left(\nt,\mt\right)$ and $\left(\nG,\mg\right)$ of the graphene and \tmd\ respectively, given by:
            \begin{align}
                \Rnm &= n \aone + m \atwo = \left(n - \frac{m}{2}\right)a\ux + \frac{m\sqrt{3}}{2}a\uy,
            \end{align}
            with $a$ the length of the unit cell.
            
            The relative strain between the layers is given by the relative length difference of the graphene and \tmd\ lattice vectors,
            \begin{align}
                \straintot = \frac{\left\vert \Rtnm\right\vert-\left\vert \Rgnm \right\vert}{\left\vert \Rgnm \right\vert }.
            \end{align}
            This $\straintot$ is applied to the graphene layer with a biaxial compression or expansion.

            The cutouts for each layer have a rotation given by $\rotlayg$ for graphene and $\rotlayt$ for \tmd compared to the reference unit cell given by \eqnref{eq:grapheneunit cell}.
            These angles are given by
            \begin{align}
                \rotlayt &= \text{arctan}\left(\frac{\mt\sqrt{3}}{2\nt - \mt}\right), &
                \rotlayg &= \text{arctan}\left( \frac{\mg\sqrt{3}}{2\nG - \mg}\right).\nonumber
            \end{align}
            The total rotation angle between the layers is given by $\rottot = \rotlayg-\rotlayt$, with $\rotlayg$ and $\rotlayt$ the rotation of each layer.
            The layers are thus rotated until the two vectors are aligned with each other.
            The angles are defined by $\left\{\rotlayg,\rotlayt,\rottot\right\}\in\left[0,2\pi\right)$.
            A summary of these transformations to construct the supercell is given in \figref{fig:supercell_construction}.
            At a twist angle of $\rottot=0\degre$, the both layers have unit cell vectors that are collinear, whereas at $30\degre$ the armchair direction of graphene runs along the zigzag direction of the \tmd.
            
        \subsection{Calculating the Low-Energy Bands}
            The total number of unit cells of graphene ($N_\gra$) or \tmd\ ($N_{\tmd}$) within the supercell characterized by $\left(\nG,\mg,\nt,\mg\right)$ is given by
            \begin{align}
                N_{\gra} &= \mg ^2 - \mg * \nG + \nG ^2, &N_{\tmd} &= \mt ^2 - \mt * \nt + \nt ^2.
            \end{align}
            For graphene, the Dirac cone is mapped to the $\Gamma$-point when $N_\gra \equiv \left(\mg+\nG\right) \mod 3$~\cite{naimer_twist-angle_2021}.
            Otherwise, the Dirac cone is mapped to the $K$ point of the supercell's Brillouin zone.
            
            In the calculations for the heterostructure, we only use the folding to $K$ and $K'$ points, as disentangling the numerical degenerate bands at the $\Gamma$ point leads to relative large numerical errors in the values for the spin expectation values compared to the well-captured values at the $K$ and $K'$-points.
            
            For each combination of $\left(\nG,\mg,\nt,\mt\right)$ we can thus obtain the twist angle $\rottot$, the strain $\straintot$ and the sizes of the supercells $N_G$ and $N_{\tmd}$.
            The total number of orbitals is given by $N_{orbitals}=4N_\gra+22N_{\tmd}$ which defines the size of the Hamiltonian which will be diagonalized to obtain the spectrum around the Dirac point. 
            
            The combination $\left(\nG,\mg,\nt,\mt\right)$ are obtained by iterating over all the different values from a certain $-n_{\max}$ to $n_{\max}$, where we selected candidate unit cells up to $n_{max}=20$.
            This search will find all possible configurations, including duplicates that have an additional rotation or that have integer multiples of the smallest possible supercell with a strain up to $|\strainmax|$.
            To obtain unique configurations, the smallest possible unit cell is chosen that has a positive angle $\rotlayg$ closest to $0\degre$.
            The number of orbitals ($N_{orbitals}$) is chosen to be less than $2000$.
            The resulting configurations for graphene have $\max\fo{\nG,\mg}=13$, and for \tmd\ $\max\fo{\nt,\mt}=9$.
            
            For the \tmd\ tight-binding model, the unstrained tight-binding parameters are used.
            This may slightly affect properties that depend on the size of the unit cell, such as the effective electron mass at certain positions in the reciprocal space.
            However, the interlayer hoppings get different lengths and bond angles depending on the applied strain on the layers.
            We use scripts adapted from~\cite{smeyers_strong_2023, andelkovic_double_2020} to construct the Hamiltonian.

            The energies in the Hamiltonian are constructed with the Dirac cone of graphene to be at zero.
            We distinguish between the energy $\engen$ (Fermi energy or energy eigenvalue, depending on the context) and the tight-binding parameters for on-site energies $\onsitetb$. In our calculations, we set the on-site energy of graphene at zero, $\onsg=0 eV$.
            However, the energy alignment of the \tmd\ changes with the band alignment, the \tmd\ on-site energies are actually shifted accordingly.
            For clarity, in the main text an arrow is used to indicate the level of the graphene on-site energy relative to the \tmd\ valence band edge, $\onsg\rightarrow\bav$.
            A practical aspect of placing the original Dirac cone at zero, is that the shift of the Dirac cone due to the proximity of the \tmd\ layer is immediately visible through the parameter $\ons$.
            
        \subsection{Flat Structure}
            The effects of in-plane and out-of-plane deformations are outside the scope of this work as this is already a non-trivial problem for monolayer \tmd~\cite{pearce_tight-binding_2016, vitale_flat_2021, fang_electronic_2018}.
            The effects of in- or out-of-plane relaxations are not taken into account.
            The main effect of relaxation on graphene's band structure can be approximated by a change in the slope of the Dirac cone, equivalent to changing $t$.
            In the \tmd\ however, the extrema of the energies at the band edges are changed, going from a direct to indirect band gap depending on the applied strain.
            Furthermore, the effect of buckling (out-of-plane deformations) on a tight-binding model for \tmd\ has hitherto never been included on a large scale TB calculation~\cite{pearce_tight-binding_2016}, whereas the effect of the bond-length variation between the layers has been used before~\cite{vitale_flat_2021}.
            Therefore, we only consider the unrelaxed structure for the TB model.
    
    \section{Effective Model}\label{sec:apx:effective_model}
            The effective graphene Hamiltonian can be expanded around the Dirac points yielding a long-wavelength continuum model, as (see \cite{sierra_van_2021,garcia_spin_2018,perkins_spintronics_2024} and references therein):
            \begin{align}
                \hat H_{G}^{eff} = &\hbar v_F \left(\tau_z\sigma_x k_x+\tau_0\sigma_y k_y\right)s_0+\oig\sigma_zs_0\tau_0\\
                & + \lkm\sigma_z s_z\tau_z \nonumber\\
                & + \lsv\sigma_0 s_z \tau_z \nonumber\\
                & + \lr e^{-i\frac{\prh s_z}{2}}\left(\tau_z\sigma_x s_y - \tau_0\sigma_y s_x\right)e^{i\frac{\prh s_z}{2}}\nonumber\\
                & + \lpi \ag \left(k_x s_y-k_y s_x\right)\sigma_z\tau_0 \nonumber\\
                & + \lpid \ag \left(k_x s_y-k_y s_x\right)\sigma_0\tau_0.\nonumber
            \end{align}
            Here, $\{\boldsymbol{s},\boldsymbol{\sigma},\boldsymbol{\tau}\}$ are Pauli matrices acting on spin ($s$), sublattice ($\sigma$), and valley ($\tau$) degrees of freedom. Note that the mass term $\oig$, Kane-Mele SOC $\lkm$ and the PIA parameters $\lpi$, $\lpid$ are set to zero in the main manuscript.
            Furthermore, the SOC energy scales can be written in terms of sublattice resolved SOC energies as:
            \begin{align}
                \lkm &= \frac{\lambda_I^A+\lambda_I^B}{2}, &
                \lsv &= \frac{\lambda_I^A-\lambda_I^B}{2},
            \end{align}
            and
            \begin{align}
                \lpi&=\frac{\lpi^A+\lpi^B}{2},&
                \lpid&=\frac{\lpi^A-\lpi^B}{2}.
            \end{align}

            \begin{figure*}
                \centering
                \includegraphics[width=\linewidth]{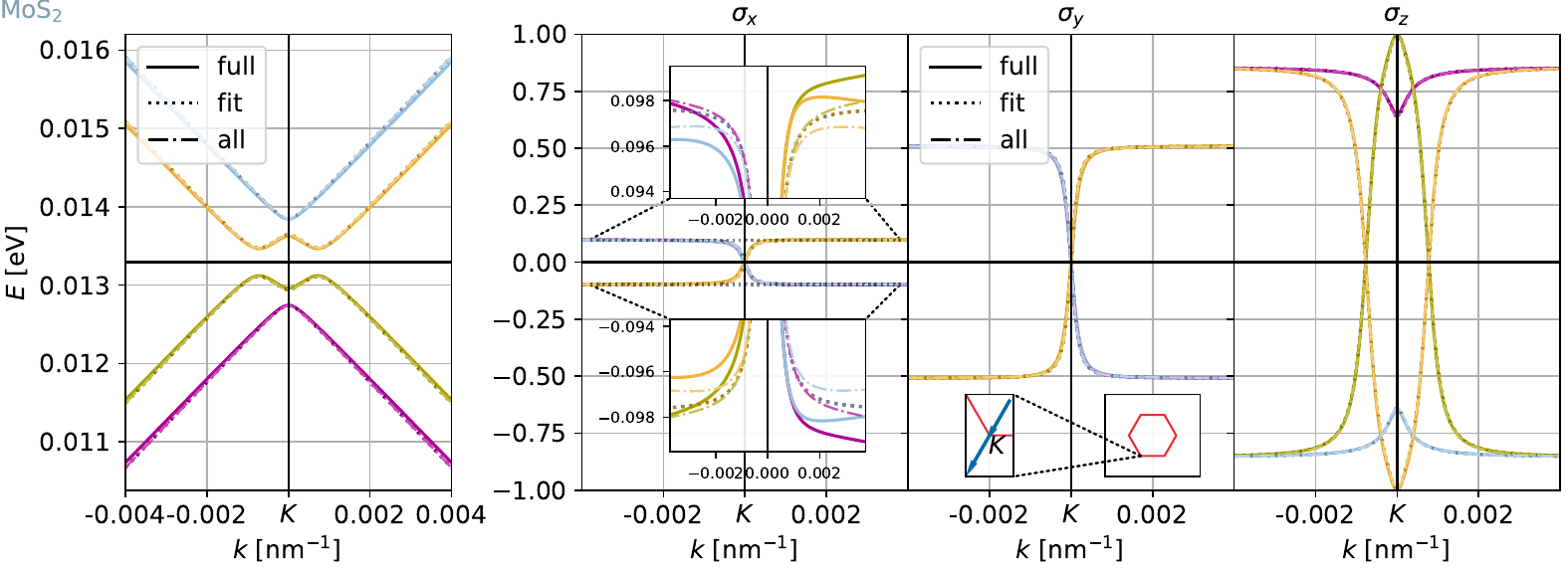}
                \caption{Similar to \figref{fig:low_energy_bands_spin}, but including $\lkm$, $\lpi$, $\lpid$ and $\oig$. (left) The low-energy bands and (right) the spin expectation values for the full (solid), effective regular (dotted) and all-parameter (dash-dotted) tight-binding model, for \mos\ with $(7,-5,4,-5)$ as shown in \figref{fig:supercell}, with $\rottot=9.17\degre$, $\straintot=-3.28\%$ and $\bac=0.04$ eV. The fitted parameters for the all-parameter model on this calculation are $\lsv=0.35$ meV, $\lr=0.21$ meV, $\prh=-1.24$ rad, $\ons=13.29$ meV, $\lpi=-0.30$ meV, $\oig=13.25$ µeV, $\lpid=-2.39$ meV and $\lkm=59.03$ neV.}
                \label{fig:apx:low_energy_bands_spin_all}
            \end{figure*}
        
            The influence of these additional parameters $\fo{\oig,\lkm,\lpi,\lpid}$ on the low-energy band structure was investigated by fitting the continuum model parameters to a representative case (see \figref{fig:apx:low_energy_bands_spin_all}).
            However, the effect of these terms on the band structure and spin expectation values is very small, which justifies the approximations made in the main text. Note that the fitted values of $\oig$ and $\lkm$ are negligible, whilst the PIA parameters are comparable to the main SOC energy scales $\lsv$ and $\lr$.
            However, we want to emphasize that these additional parameters are useful in first-principles studies. There, the structure is often relaxed first~\cite{venkateswarlu_electronic_2021}.
            This relaxation can generate a significant sublattice imbalance, and thus a more pronounced $\oig$ and $\lkm$.

            We want to note that the variables used by \lk~\cite{li_twist-angle_2019} have a different pre-factor.
            In our convention, the band gap is $2\lsv$ at the Dirac point, the splitting of the graphene energy cone away from Dirac cone is given by $2\sqrt{\lsv^2+\lr^2}$, as seen on \figref{fig:low_energy_bands_spin} by the splitting of the blue/yellow and green/pink bands.
            The data from \lk\ given in \figref{fig:parameters_angle} have been scaled accordingly. 

    \section{Fitting Routine}\label{sec:apx:fitting_routine}
        The parameters are fitted to reproduce the full TB calculations in several stages, first a fit of the effective model, then a harmonic fit across twist angles and finally a polynomial fit across realisations of the band alignment or charge-transfer dipole.

        \subsection{Calculation of the Full Tight-Binding Model}
            The full tight-binding model is solved by exact diagonalization using the Arnoldi method implemented in SciPy~\cite{virtanen_scipy_2020}.
            The energy bands are calculated across a cut that traverses the $K$-point of the mini-Brillouin-zone.
            The orbital contribution of graphene to these energy bands is determined, from which we select the four energy bands with the highest graphene orbital contribution.
            These bands are then analysed, where they are disentangled by taking the largest overlap in eigenvector for the eigenvalues along the cut.
            The $x$-, $y$- and $z$-spin expectation values are extracted from the associated eigenvectors for the energy bands.
            
        \subsection{Fit to the Full Tight-Binding Model}
            Each configuration of the full TB calculation is used as a basis for fitting the effective TB model.
            The fit aims to yield a correct description of the 4 low-energy bands around the Dirac points and the spin expectation values for these bands.
            To this end, the sum of the squared differences between the energies and spin expectation values of the two models is minimized.
            The spin expectation values have a weight factor $w^s$, with $w_x^s=w_y^s=0.006$ and $w_z^s=0.003$.
            The total cost function is then given as
            \begin{align}
                \chi^2 &= \sum_{i=1}^{n_k} \sum_{j=1}^4 \left[\left(E_{i,j}^{full}-E_{i,j}^{eff}\right)^2+\sum_{l=\{x,y,z\}}w^{s}_l\left(s_l^{full}-s_l^{eff}\right)^2\right],\nonumber
            \end{align}
            where $i,j,l$ run over reciprocal space points, energy bands and spin directions, respectively.
            The resulting cost function $\chi^2$ is minimized using the \textit{minimize} function of SciPy~\cite{virtanen_scipy_2020}.

        \subsection{Harmonic Fit}
            The effective model parameters are extracted for various twist angles and fitted to a minimal Fourier series to determine their $\rottot$ dependence ~\cite{perkins_spin_2024}, namely:
            \begin{align}\label{eq:fitfunc}
                \fitharlsv\left(\rottot\right) &= \sum_{i=0}^3 \fitharparlsv{i} \cos\left(2\pi \frac{\rottot}{120\degre} (1+2i) \right), \\
                \fitharlra\left(\rottot\right) &= \sum_{i=0}^3 \fitharparlra{i} \cos\left(2\pi \frac{\rottot}{60 \degre}  i     \right), \nonumber \\
                \fitharons\left(\rottot\right) &= \sum_{i=0}^3 \fitharparons{i} \cos\left(2\pi \frac{\rottot}{60 \degre}  i     \right), \nonumber \\
                \fitharcpr\left(\rottot\right) &= \sum_{i=0}^3 \fitharparcpr{i} \cos\left(2\pi \frac{\rottot}{60 \degre}  i     \right), \nonumber \\
                \fitharspr\left(\rottot\right) &= \sum_{i=0}^3 \fitharparspr{i} \sin\left(2\pi \frac{\rottot}{60 \degre}  (1+i) \right), \nonumber \\
                \fitharprh\left(\rottot\right) &= \arctan_2\left(\fitharspr\left(\rottot\right), \fitharcpr\left(\rottot\right)\right), \nonumber &
            \end{align}
            with $\arctan_2\left(y,x\right)$ the inverse $\tan$-function as implemented in NumPy~\cite{harris_array_2020}.
            Note that $\fitharlsv$ vanishes for $\rottot=30\degre + n \, 60 \degre$ owing to the existence of a vertical mirror plane of symmetry at those twist angles.
            \begin{table}
                \centering
                \begin{tabular}{rl|rrrr}
                                    &           & $\fitharpargen{0}$    & $\fitharpargen{1}$    & $\fitharpargen{2}$    & $\fitharpargen{3}$    \\ \hline
                    $\fitharlsv$    & $[meV]$   &  0.443& -0.218&  0.002&  0.049 \\
                    $\fitharlra$    & $[meV]$   &  0.159& -0.006& -0.069&  0.007 \\
                    $\fitharcpr$    & $[/]$     &  0.251&  0.259& -0.086&  0.386 \\
                    $\fitharspr$    & $[/]$     & -1.198& -0.046& -0.328&  0.047 \\
                    $\fitharons$    & $[meV]$   & 19.918& -8.786&  3.858& -0.836 \\
                \end{tabular}
                \caption{The extracted Fourier coefficients for \mos\ with no dipole correction applied and $\bacz=0.04$ eV.}
                \label{tab:fitcoefmos}
            \end{table}
            Representative Fourier coefficients are provided in \tabref{tab:fitcoefmos}.

        \subsection{Polynomial Fit}\label{sec:apx:polynomial_fit_ba_dipole}
            \begin{table*}
                \centering
                \begin{tabular}{rl|rrrrrrr}
                    & & $\fitpolpargen{j}{0}$ & $\fitpolpargen{j}{1}$ & $\fitpolpargen{j}{2}$ & $\fitpolpargen{j}{3}$ & $\fitpolpargen{j}{4}$ & $\fitpolpargen{j}{5}$ & $\fitpolpargen{j}{6}$ \\ \hline
                    $\fitpollsv{0}$ & $[meV]$   &  0.470& -0.741&  0.858& -0.306& -0.387&  0.408& -0.115 \\
                    $\fitpollsv{1}$ & $[meV]$   & -0.250&  0.783& -1.093&  0.444&  0.533& -0.576&  0.161 \\
                    $\fitpollsv{2}$ & $[meV]$   &  0.006& -0.121&  0.195& -0.110& -0.106&  0.135& -0.044 \\
                    $\fitpollsv{3}$ & $[meV]$   &  0.052& -0.109&  0.127& -0.022& -0.043&  0.027&  0.000 \\
                    $\fitpollra{0}$ & $[meV]$   &  0.167& -0.186&  0.229&  0.019& -0.063& -0.002&  0.018 \\
                    $\fitpollra{1}$ & $[meV]$   & -0.006&  0.016& -0.011& -0.086& -0.017&  0.105& -0.049 \\
                    $\fitpollra{2}$ & $[meV]$   & -0.068& -0.011&  0.417& -1.201&  1.516& -0.884&  0.195 \\
                    $\fitpollra{3}$ & $[meV]$   &  0.010& -0.056&  0.163& -0.099& -0.084&  0.120& -0.039 \\
                    $\fitpolcpr{0}$ & $[/]$     &  0.227&  0.828&  1.211& -3.691&  3.702& -1.694&  0.295 \\
                    $\fitpolcpr{1}$ & $[/]$     &  0.305& -1.136& -1.795&  8.247& -9.511&  4.646& -0.839 \\
                    $\fitpolcpr{2}$ & $[/]$     & -0.081& -0.113&  3.470& -8.912&  9.170& -4.272&  0.749 \\
                    $\fitpolcpr{3}$ & $[/]$     &  0.400& -0.522& -2.268&  7.516& -8.559&  4.276& -0.791 \\
                    $\fitpolspr{0}$ & $[/]$     & -1.201&  0.242&  0.831& -0.756&  0.452& -0.202&  0.045 \\
                    $\fitpolspr{1}$ & $[/]$     & -0.017& -0.695&  2.721& -6.318&  6.950& -3.453&  0.633 \\
                    $\fitpolspr{2}$ & $[/]$     & -0.315& -0.079&  0.839& -1.217&  0.841& -0.295&  0.044 \\
                    $\fitpolspr{3}$ & $[/]$     &  0.038& -0.035&  0.696& -2.409&  2.950& -1.531&  0.288 \\
                    $\fitpolons{0}$ & $[meV]$   & 19.166& 18.006&  0.449& -7.910& 15.270& -9.441&  2.365 \\
                    $\fitpolons{1}$ & $[meV]$   & -8.715& -0.321&-12.772& 30.721&-39.890& 23.576& -5.518 \\
                    $\fitpolons{2}$ & $[meV]$   &  3.983& -3.387&  9.994&-18.349& 22.293&-13.733&  3.374 \\
                    $\fitpolons{3}$ & $[meV]$   & -1.020&  5.016&-24.083& 53.343&-59.552& 32.325& -6.836 \\
                \end{tabular}
                \caption{Band-alignment expansion coefficients for \mos.}
                \label{tab:fitpolucoef}
            \end{table*}
            The variation of Fourier coefficients $\fitharpargen{i}$ with the band alignment is investigated and fitted to a power series of the form:
            \begin{align}
                \fitpolgen{j}\fo{x} &= \sum_{i=0}^6 x^i \fitpolpargen{j}{i},
            \end{align}
            representative coefficients are reported in \tabref{tab:fitpolucoef}. 

    \section{Additional Results}\label{sec:apx:more_results}
        In this section, we analyse the effects of a dipole correction and band alignment changes.
        
        \subsection{Variation of Charge-Transfer Dipole}
             \begin{figure}
                 \centering
                 \includegraphics[width=\linewidth]{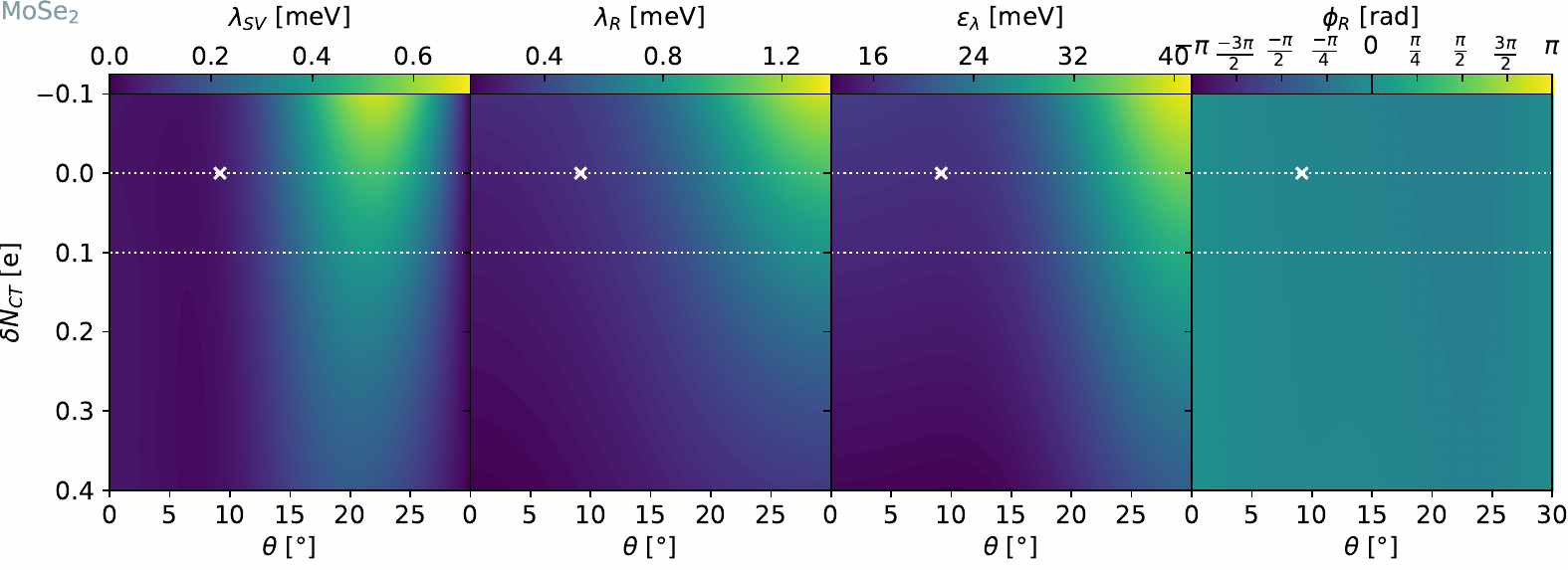}
                 \includegraphics[width=\linewidth]{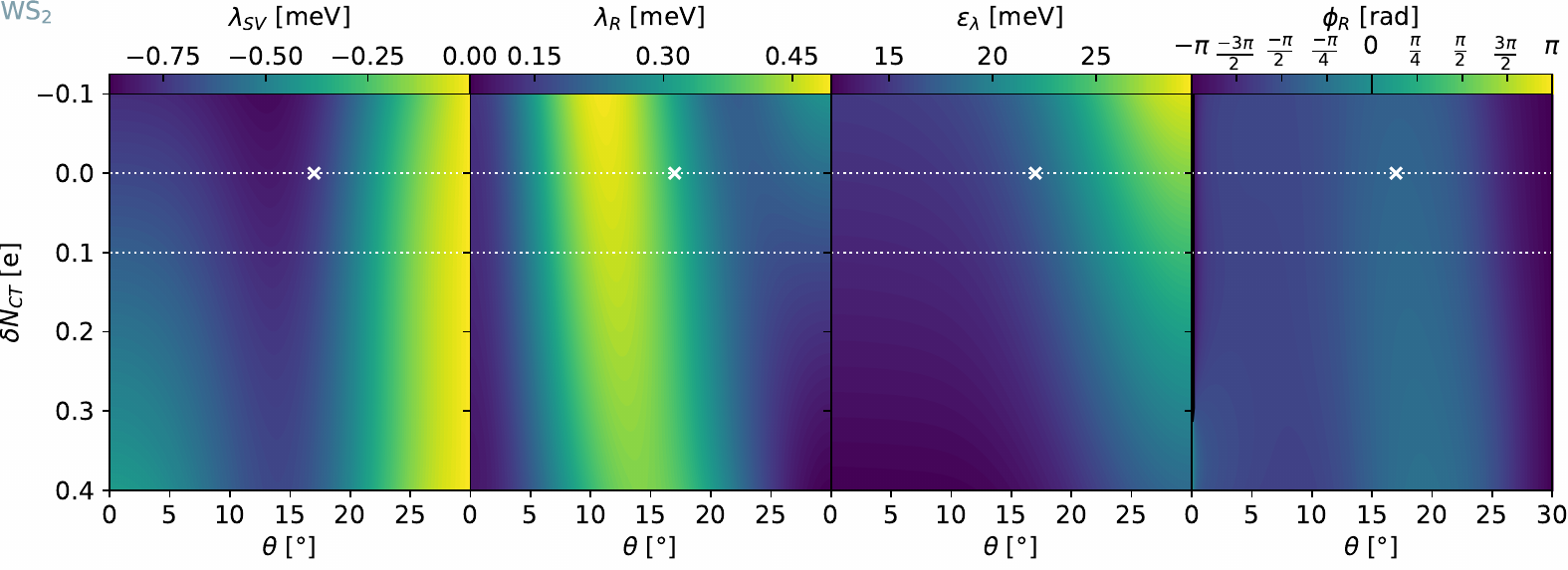}
                 \includegraphics[width=\linewidth]{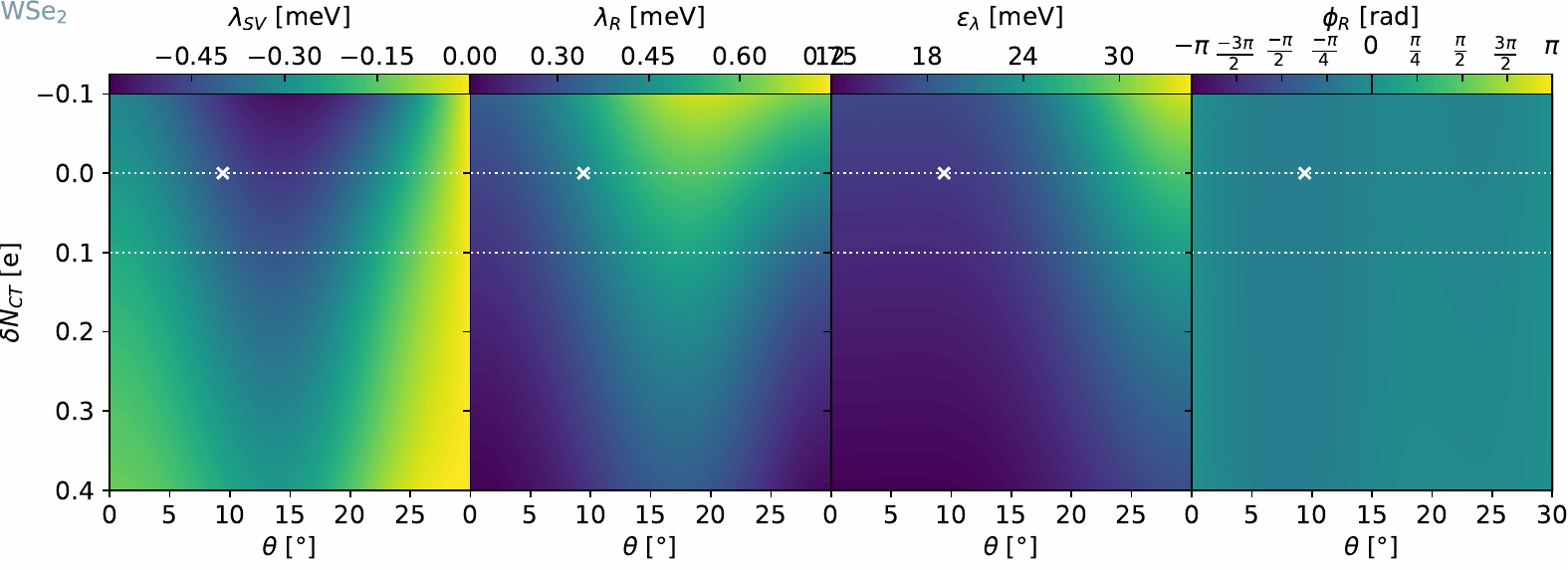}
                 \caption{Dependence of SOC with the dipole correction for (top) \mose, (middle) \ws\ and (bottom) \wse.
                 The reference band alignment $\bacz$ in \tabref{tab:interlayer_parameters} was employed.
                 Dotted lines indicate $\dct=0e$ and $+.1 e$.}
                 \label{fig:apx:dipole_more}
             \end{figure}
            The variation of the SOC parameters for \mose, \ws\ and \wse\ upon considering a charge-transfer dipole is shown in \figref{fig:apx:dipole_more}.
            As discussed in the main text, see also \figref{fig:dipole}, the introduction of a charge-transfer dipole generally produces relatively small changes in the SOC energies and Rashba phase $\prh$.
            However, there are important exceptions to this general trend.
            For example, the Rashba SOC energy $\lr$ in twisted graphene/\mose\ with $30\degre\gtrsim\rottot\gtrsim20\degre$ is seen to change by more than $60\%$ upon varying $\dct$ from $-0.1$e to $+0.1$e.
            Similar variation is seen in the Rashba coupling induced by other TMDs.
            As for the spin-valley coupling, $\lsv$, the greatest variation is also found for \mose/graphene with a change of more than $70\%$ upon varying $\dct$ from $-0.1$e to $+0.1$e.
            On the other hand, the \rashbap\ shows minute changes with the application of a charge dipole.
            For \ws/graphene, a switch from $\prh=-\pi$ to $0$ occurs with an increase in filling around $\dct$ of $+.3$e.
            
        \subsection{Variation of Band Alignment}
            \begin{figure}
                 \centering
                 \includegraphics[width=\linewidth]{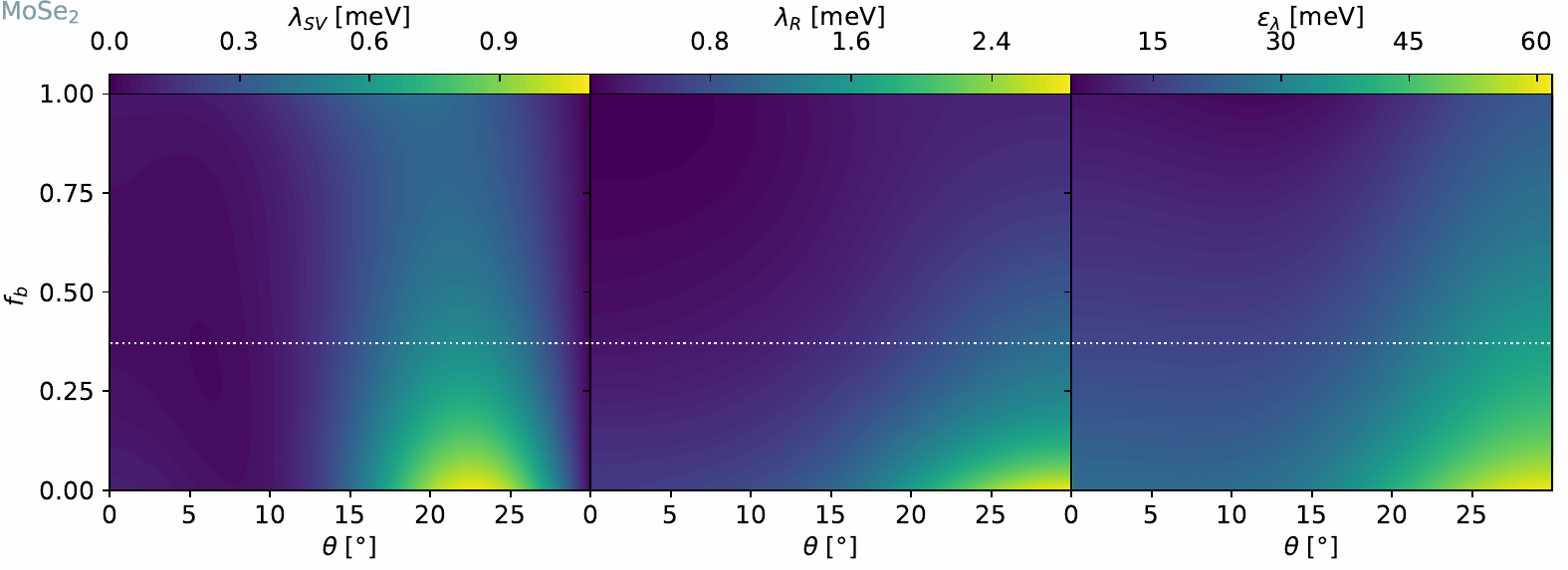}
                 \includegraphics[width=\linewidth]{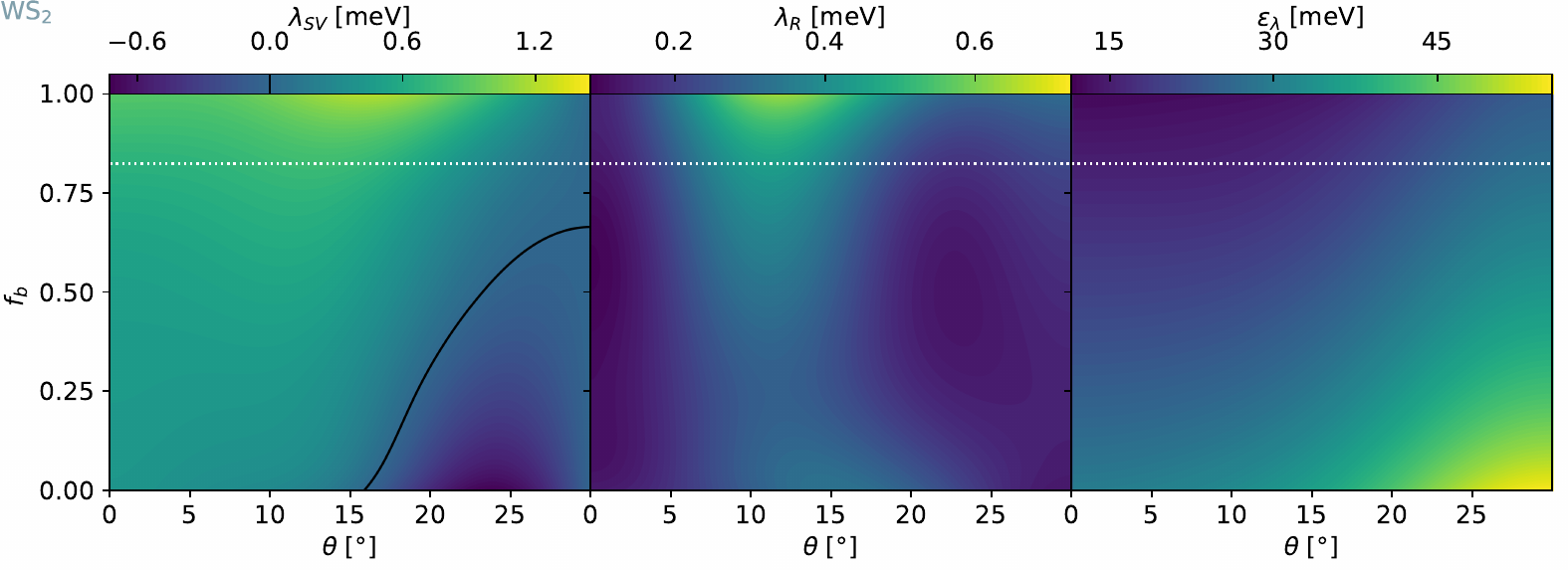}
                 \includegraphics[width=\linewidth]{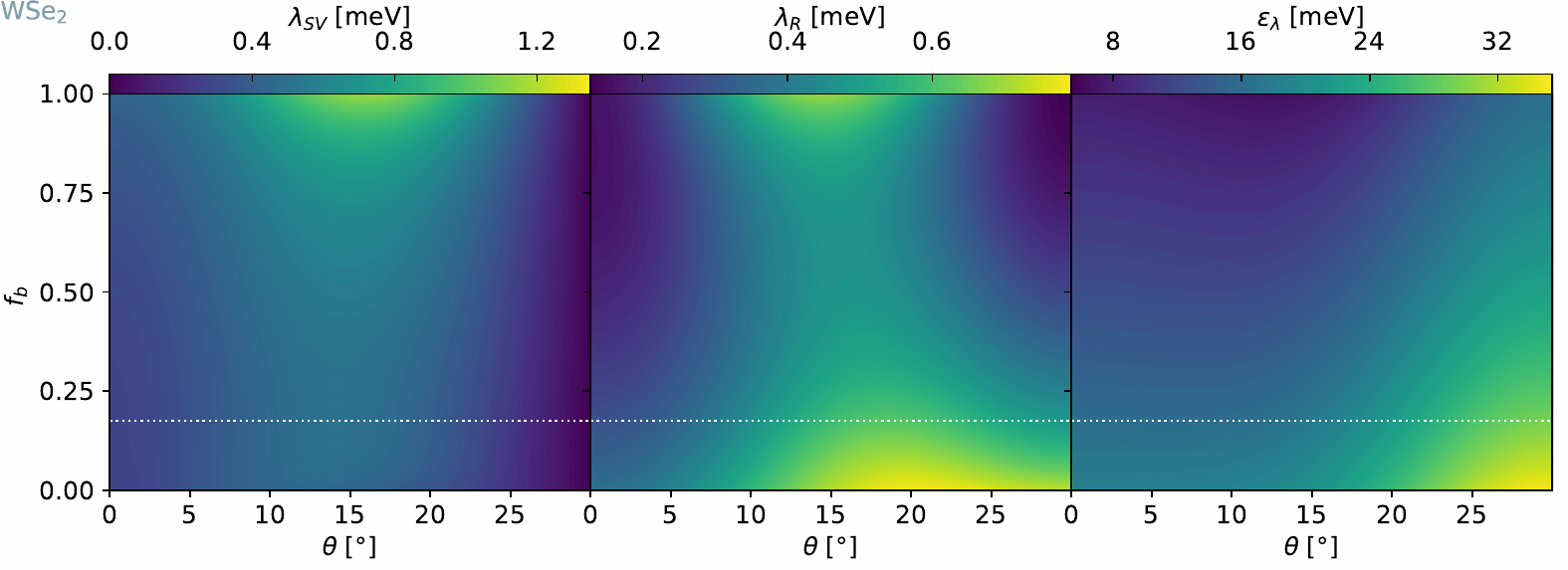}
                 \caption{Dependence of SOC with the band alignment for (top) \mose, (middle) \ws\ and (bottom) \wse.
                 No dipole correction applied. The dotted lines represent the reference fractional band alignment $\bafz$ given in \tabref{tab:interlayer_parameters}.}
                 \label{fig:apx:ba_more}
             \end{figure}
            The impact of changes in the band alignment $\baf$ is shown in \figref{fig:apx:ba_more}.
            The induced variations in SOC in twisted heterostructures are generally pronounced.
            Note the change of sign in $\lsv$ around $\baf\approx 0.65$ for graphene/\ws\ with $\rottot\approx 30\degre$.
            The transition from positive to negative $\lsv$ is pushed towards lower $\baf$ values as the twist angle approaches $\approx 15\degre$ from above.
            Below this critical twist angle, $\lsv$ becomes little sensitive to the band alignment.
            Such a change of sign of $\lsv$ is not seen in \mos\ nor in \wse\ cases.
            The Rashba SOC is also greatly renormalized by $\baf$, especially for graphene/\mose, where the $\lr$ increases for over $20\%$ going from $\baf=.75$ to $.89$, and similarly for graphene/\wse, where $\lr$ also increases in $20\%$ going from $\baf=.25$ to $\baf=.11$.
            
        \subsection{Variation of Band Alignment with Dipole}\label{sec:apx:ba_dipole}
            \begin{figure}
                 \centering
                 \includegraphics[width=\linewidth]{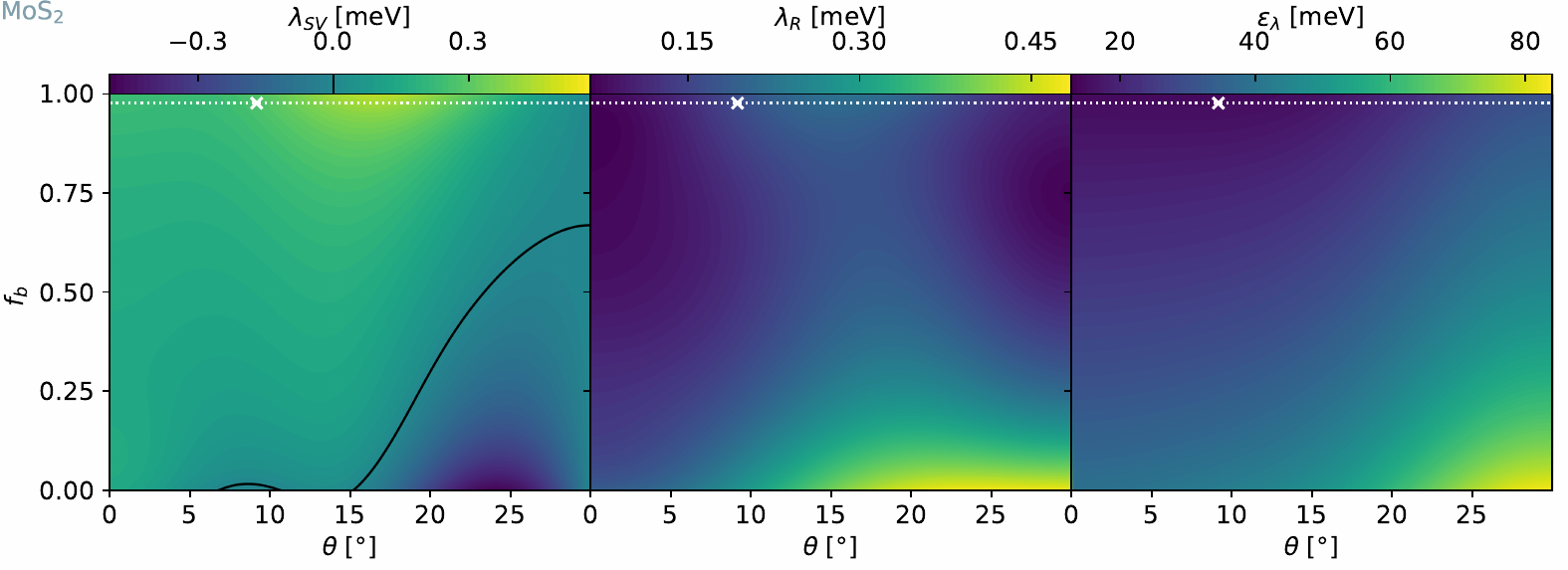}
                 \includegraphics[width=\linewidth]{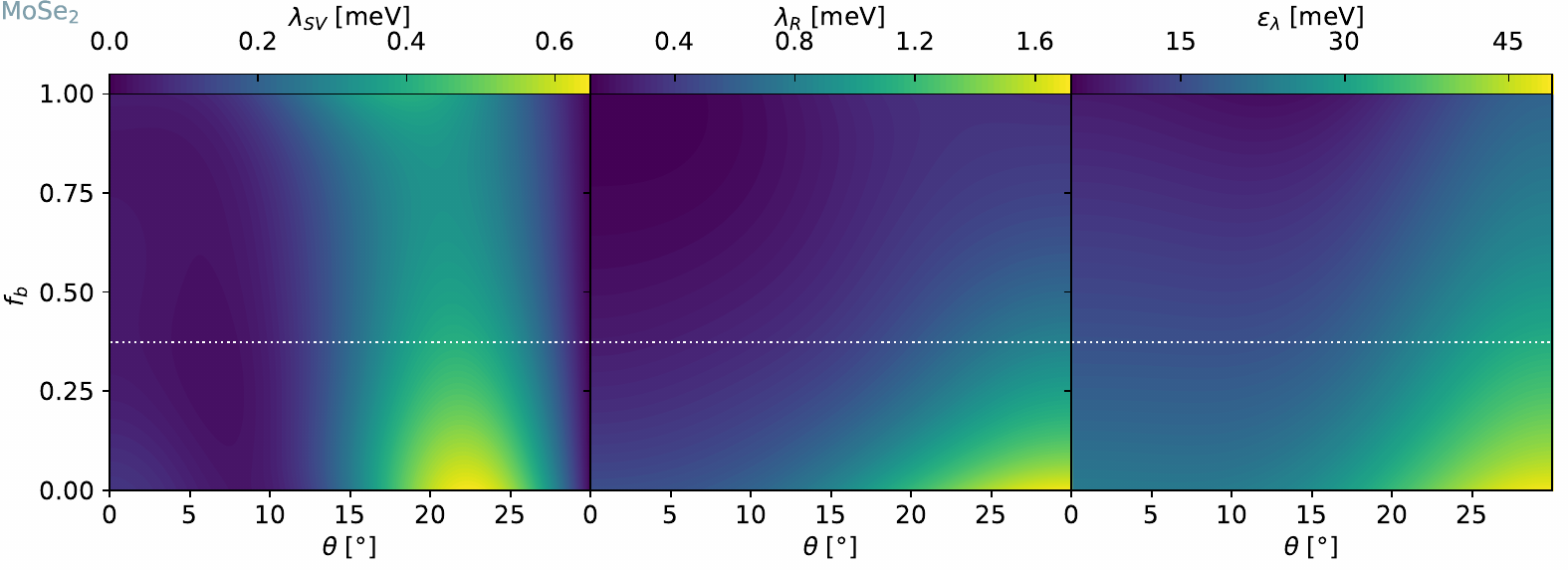}
                 \includegraphics[width=\linewidth]{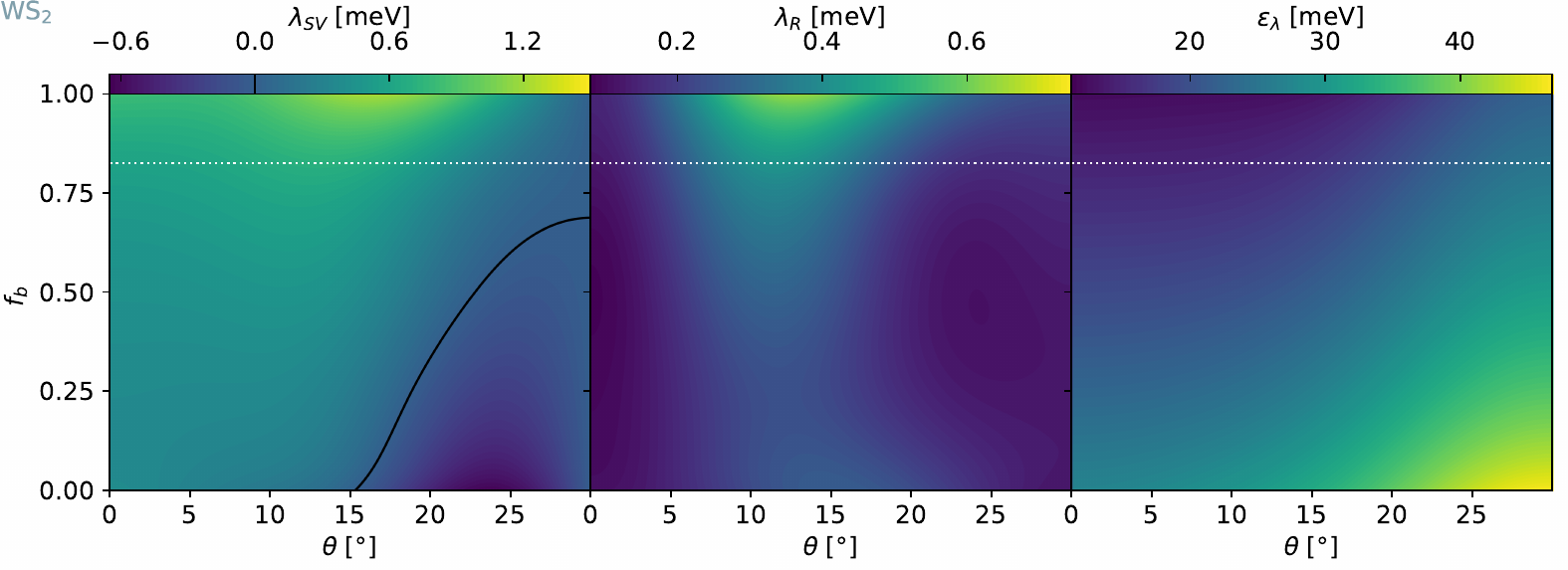}
                 \includegraphics[width=\linewidth]{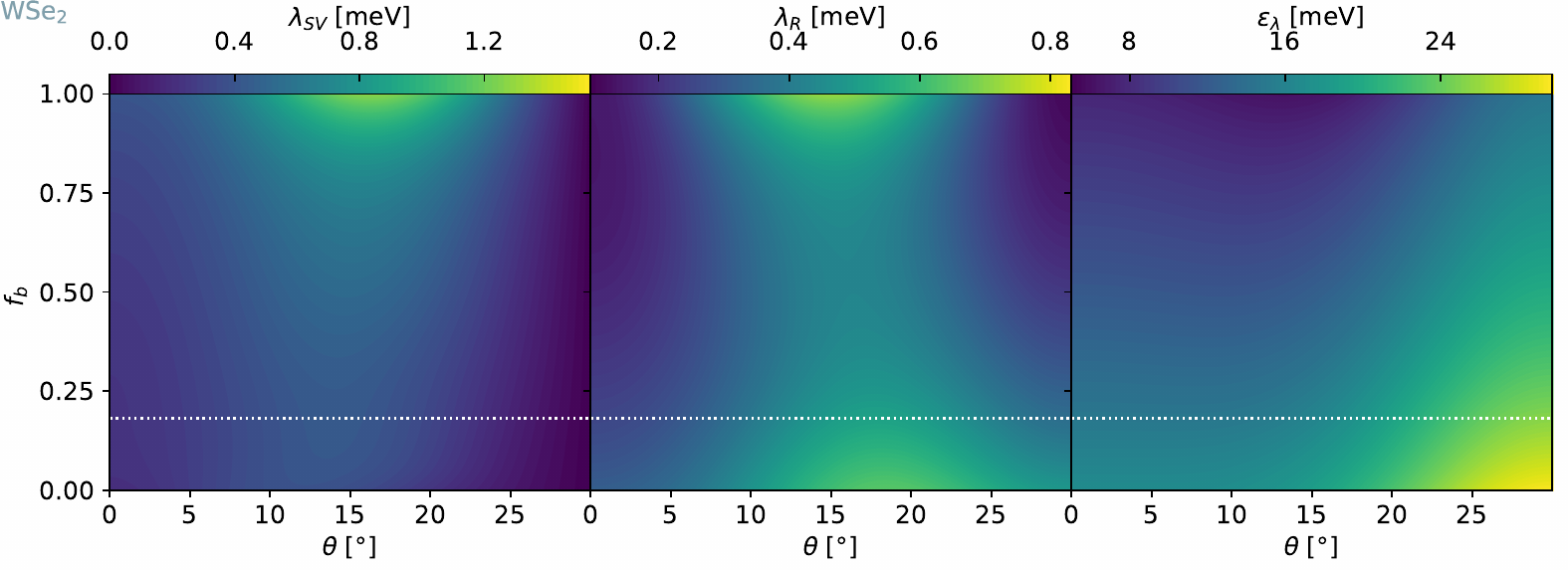}
                 \caption{Dependence of SOC with band alignment for \mos, \mose, \ws\ and \wse.
                 A dipole of $\dct=+.1e$ is applied. The dotted lines represent the reference fractional band alignment $\bafz$ given in \tabref{tab:interlayer_parameters}.}
                 \label{fig:apx:ba__dipole}
             \end{figure}
            \begin{figure}
                 \centering
                 \includegraphics[width=\linewidth]{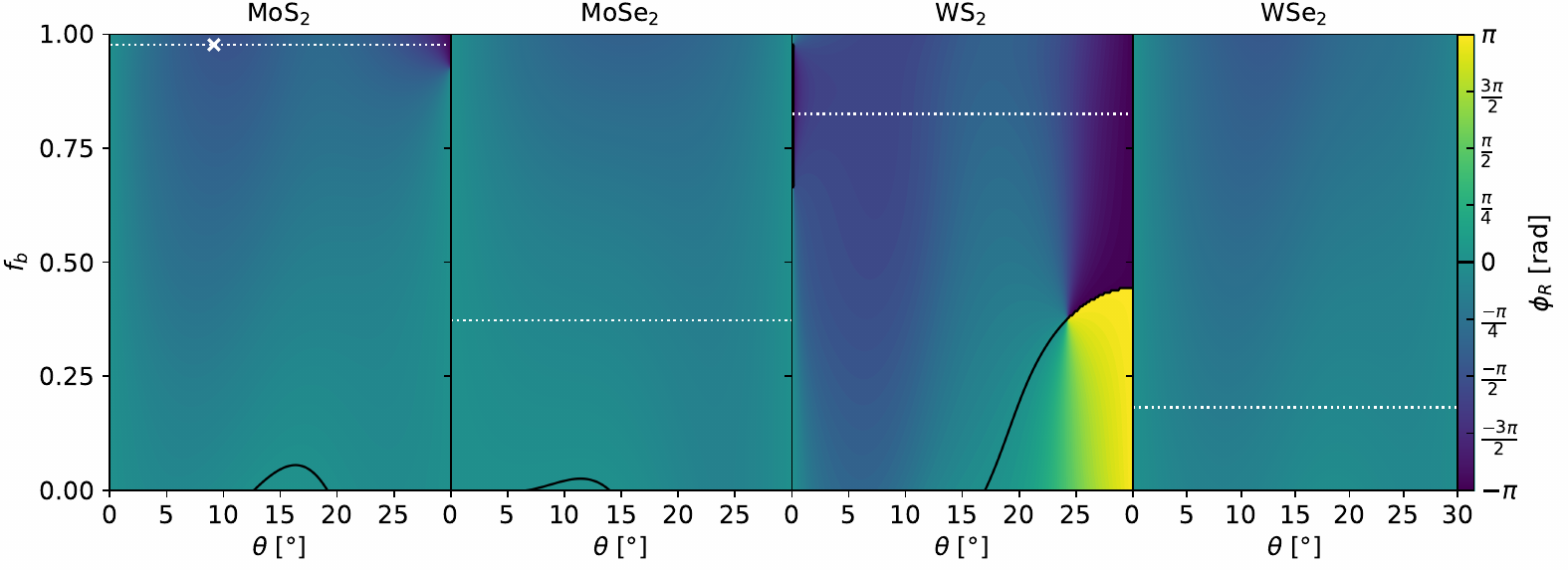}
                 \caption{Dependence of $\prh$ upon the twist angle and band alignment $\baf$ for \mos, \mose, \ws\ and \wse\ with a dipole of $\dct=+.1e$.}
                 \label{fig:apx:ba__dipole_prh}
             \end{figure}
            The band-alignment dependence for a dipole of $\dct=+.1e$ is shown in \figref{fig:apx:ba__dipole} and \figref{fig:apx:ba__dipole_prh}.
            As noted in the previous section comparing the dipole variation across the materials, the results vary only minimal with the inclusion of the dipole correction.
            
    \section{Determining the Dipole Correction}\label{sec:apx:dipole}
        \begin{figure}
            \centering
            \includegraphics[width=\linewidth]{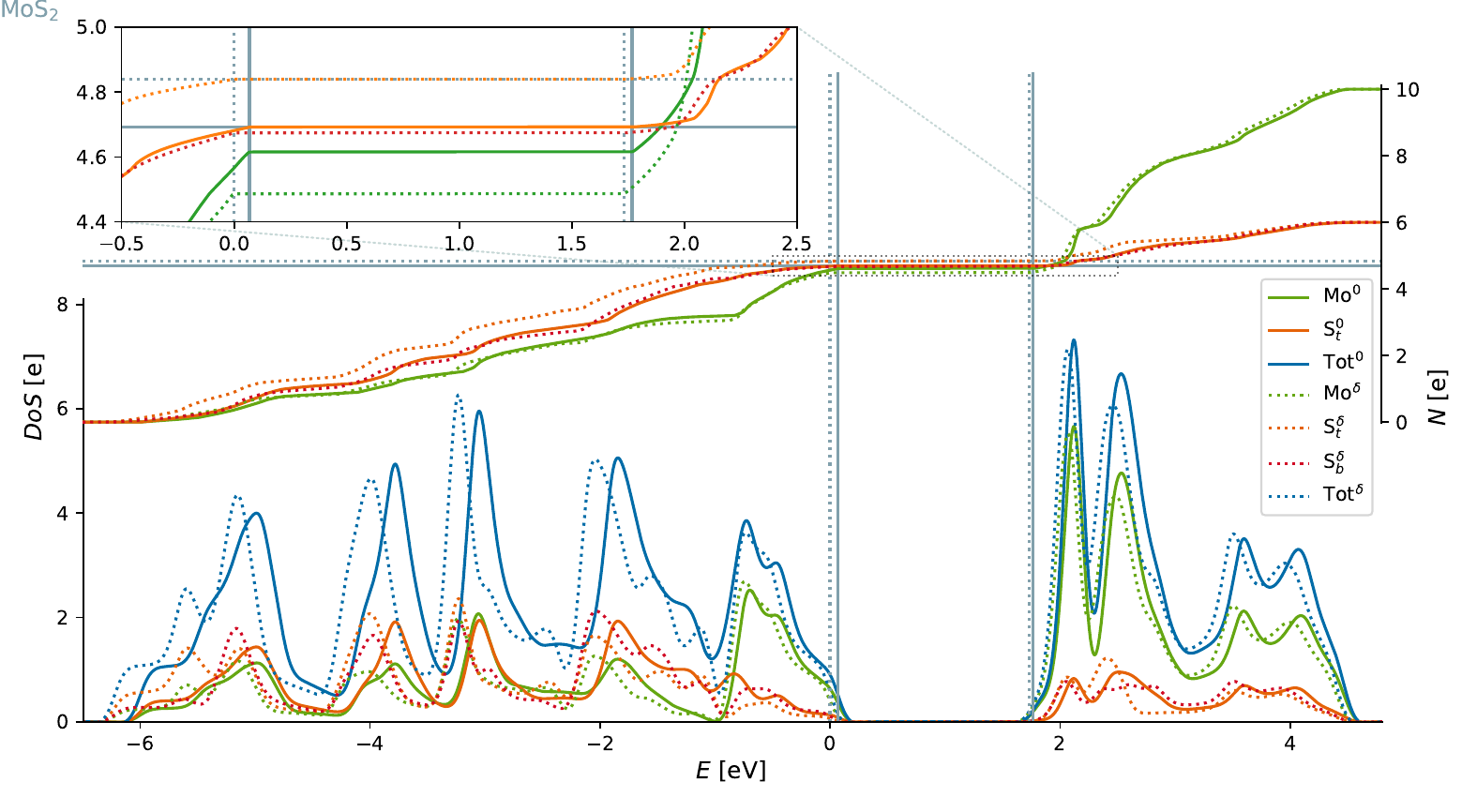}
            \caption{The \dos\ and the corresponding filling values for the different energy levels of \mos\ for the reference structure $\oxtct=0 eV$, indicated in solid with $X^0$, and with an on-site shift of $\oxtct=-0.5eV$, indicated in dotted lines with $X^\delta$, for the orbitals on the metal ($Mo$) and top ($S_t$) and bottom chalcogen ($S_b$) atoms.
            The \dos\ was smoothened with a spline for better visibility on the figure.
            For the reference levels, the \dos\ for $S_t^0$ are degenerate with $S_b^0$.}
            \label{fig:apx:dos}
        \end{figure}
        The dipole is added to the \tmd\ by changing the on-site energy for the chalcogen atoms closest to the graphene ($\tsxu$).
        This results in a shift in the filling of these orbitals, $\dct$.
        We look at the effect of filling this orbital with the change in this on-site energy $\oxtct$ to determine the on-site energy shift corresponding to a given relative filling.
    
        First, we consider the average density of states (\dos) and the projected \dos\ for the different atoms.
        From this, we can determine the accumulated filling per energy shown in \figref{fig:apx:dos}.
        The total filling of the top chalcogen orbital is 6, corresponding to the 3 orbitals and 2 electrons per orbital (spin up and down).
        The band gap is here also clearly visible as a flat line.

        Next, we examine the influence of the dipole on the filling of the $S$ orbital on which the dipole is added, which is displayed in \figref{fig:apx:dos} with dotted lines.
        This is done by changing the on-site energy of the highest chalcogen atom,
        \begin{equation}
            \oxt = \oxtz + \oxtct\fo{\dct},
        \end{equation}
        where $\oxtz$ is the on-site energy as given in the normal model~\cite{fang_ab_2015,fang_electronic_2018} and $\oxtct$ the additional on-site term shift.
        For the calculations for the \dos\ at different on-site shifts, $\oxt$ is varied.
        The aim is to correlate the used dipole correction in the on-site term $\oxtct$ to generate a corresponding filling $\dct\fo{\oxtct}$.
        Then, the filling relative to the on-site shift can be inverted to obtain $\oxtct\fo{\dct}$, the on-site shift needed to achieve a specific filling.
        
        The filling at the band gap is used to determine the total orbital filling.
        The band gap is calculated by the distance between the edges of the conduction and valance band.
        This differs from the spin degenerate case, where the splitting due to the SOC is most noticeable with the heavy metal \tmd.
        The spin degenerate band gap is $\bgt=$ 1.77 eV (\mos), 1.50 eV (\mose), 1.97 eV (\ws) and 1.66 eV (\wse).
        \begin{figure}
            \centering
            \includegraphics[width=\linewidth]{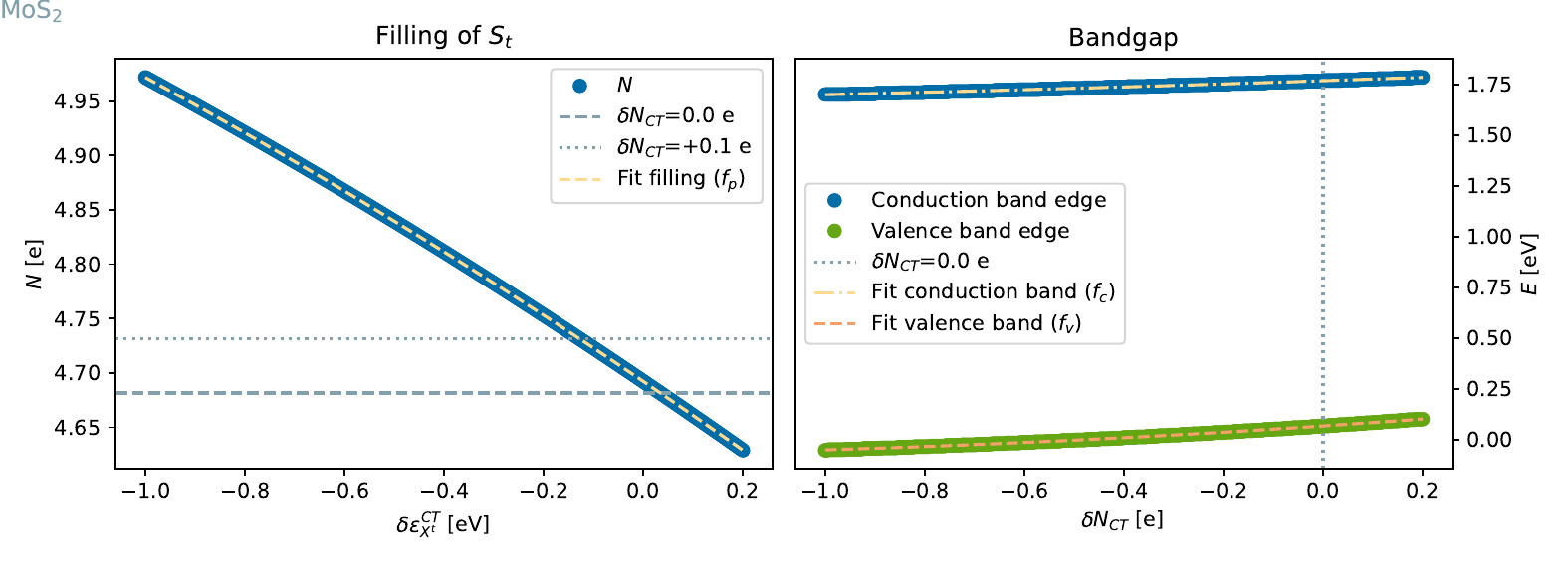}
            \caption{Filling of the $\tsxu$ $p$-orbitals with fit $f_p$ (left) and change in band edges with fits $f_v$ and $f_c$ for the valance and conduction band edges, respectively (right).}
            \label{fig:apx:filling_fit}
        \end{figure}
        In \figref{fig:apx:filling_fit}, the filling of the top chalcogen orbital with the shift and the total of all the orbitals is shown, with a corresponding quadratic fit.
        \begin{table}
            \centering
            \begin{tabular}{lr|rlrlrl}
                        &       & $a$       &               & $b$       &           & $c$       &           \\ \hline
                \mos    & $f_p$ & -0.0305   & $[e/eV^2]$    & -0.3102   & $[e/eV]$  &  4.6922   & $[e]$     \\
                        & $f_v$ &  0.0446   & $[eV/e^2]$    &  0.1621   & $[eV/e]$  &  0.0672   & $[eV]$    \\
                        & $f_c$ &  0.0122   & $[eV/e^2]$    & 0.0810    & $[eV/e]$  &  1.7677   & $[eV]$    \\
                \mose   & $f_p$ & -0.0347   & $[e/eV^2]$    & -0.3542   & $[e/eV]$  &  4.6047   & $[e]$     \\
                        & $f_v$ &  0.0094   & $[eV/e^2]$    &  0.0904   & $[eV/e]$  &  0.0362   & $[eV]$    \\
                        & $f_c$ &  0.0119   & $[eV/e^2]$    &  0.0753   & $[eV/e]$  &  1.5004   & $[eV]$    \\
                \ws     & $f_p$ & -0.0256   & $[e/eV^2]$    & -0.2644   & $[e/eV]$  &  4.8176   & $[e]$     \\
                        & $f_v$ &  0.0077   & $[eV/e^2]$    &  0.0982   & $[eV/e]$  &  0.2697   & $[eV]$    \\
                        & $f_c$ &  0.0056   & $[eV/e^2]$    &  0.0662   & $[eV/e]$  &  1.9723   & $[eV]$    \\
                \wse    & $f_p$ & -0.0296   & $[e/eV^2]$    & -0.3056   & $[e/eV]$  &  4.7297   & $[e]$     \\
                        & $f_v$ &  0.0077   & $[eV/e^2]$    &  0.0931   & $[eV/e]$  &  0.2693   & $[eV]$    \\
                        & $f_c$ & -0.0314   & $[eV/e^2]$    &  0.0488   & $[eV/e]$  &  1.6638   & $[eV]$
            \end{tabular}
            \caption{Dipole correction filling-coefficients, with $f_p$ the fit for the filling $N$ of the $\tsxu$ orbital in function of the on-site term shift $\oxtct$ as indicated on the left pane in \figref{fig:apx:filling_fit}, $f_v$ and $f_c$ indicate the coefficients for the fit of the band edges in function of the additional filling $\dct$, as indicated on the right pane of the same figure.}
            \label{tab:apx:filling}
        \end{table}
        The change in the band gap with the application of the dipole is shown in \figref{fig:apx:filling_fit}.
        The band edges only change slightly with the application of the dipole.

        The filling ($\dct$) in units of electrons transferred is thus given as a function of the parameters ($\oxtct$),
        \begin{align}
            \dct = \fo{\oxtct}^2a+\fo{\oxtct} b+c
        \end{align}which is transformed to
        \begin{align}
            \oxtct = \frac{-b-\sqrt{b^2-4a(-\dct)}}{2a},
        \end{align}
        where the formula is simplified such that the filling at neutrality (no dipole) is taken as zero filling, with the coefficients $a$, $b$ and $c$ given in \tabref{tab:apx:filling}.
        This value for the on-site shift $\oxtct\fo{\dct}$ is then used in the TB model for the top chalcogen atom.
        The fits for the band edges given in \figref{fig:apx:filling_fit} and \tabref{tab:apx:filling} yield the filling dependence of the band gap.
            The changes in fractional band alignment $\baf$ are correlated with this changing band gap.
                
    \section{Remarks on the Rashba Phase}\label{sec:apx:phir_ws}
        \begin{figure}
            \centering
            \includegraphics[width=\linewidth]{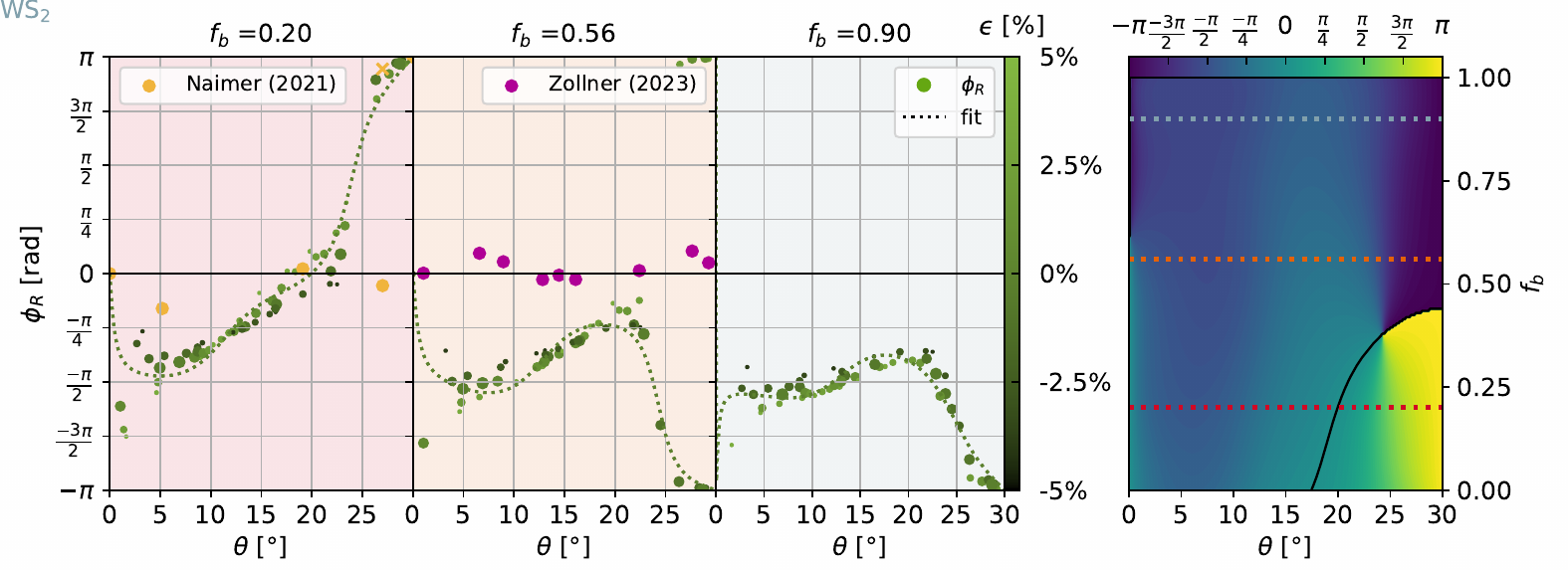}
            \caption{(left) Fits for $\prh\fo\rottot$ and comparison with the literature for \ws\ at selected values of $\baf$. (right) Density plot of Rashba phase in the $(\rottot,\baf)$-parameter space.}
            \label{fig:apx:phir_ws2}
        \end{figure}
            The non-trivial behaviour of the \rashbap\ for \ws\ in the parameter space of the TB model can be appreciated in \figref{fig:ba_phi} (main text), where the solid line marks a branch cut singularity. As the cut is traversed, a Rashba angle $\pm \pi$ rad jumps to $\mp \pi$ rad.
            In the results from \textit{Naimer et al.}\cite{naimer_twist-angle_2021}, given in their table VII, together with those indicated in their Fig. 7, there seems to be an inconsistency for the \ws\ results at $30\degre$, where the table reports $+\pi$ rad, and the figure indicates $0$ rad.
            Assuming that the table is in fact correct, their data for $27\degre$, $-0.19$ rad can be shifted similarly with an addition of $+\pi$ rad as was done at $30\degre$.
            The five data points, with the two points at $27\degre$ and $30\degre$ both uncorrected (solid circle) and corrected (cross) with a shift of $+\pi$ rad are shown in \figref{fig:apx:phir_ws2}, which overall agree qualitatively well with our results.
            We also included the results from \textit{Zollner et al.}, but these seem to deviate from our results considerably.

            \begin{figure}
                \centering
                \includegraphics[width=\linewidth]{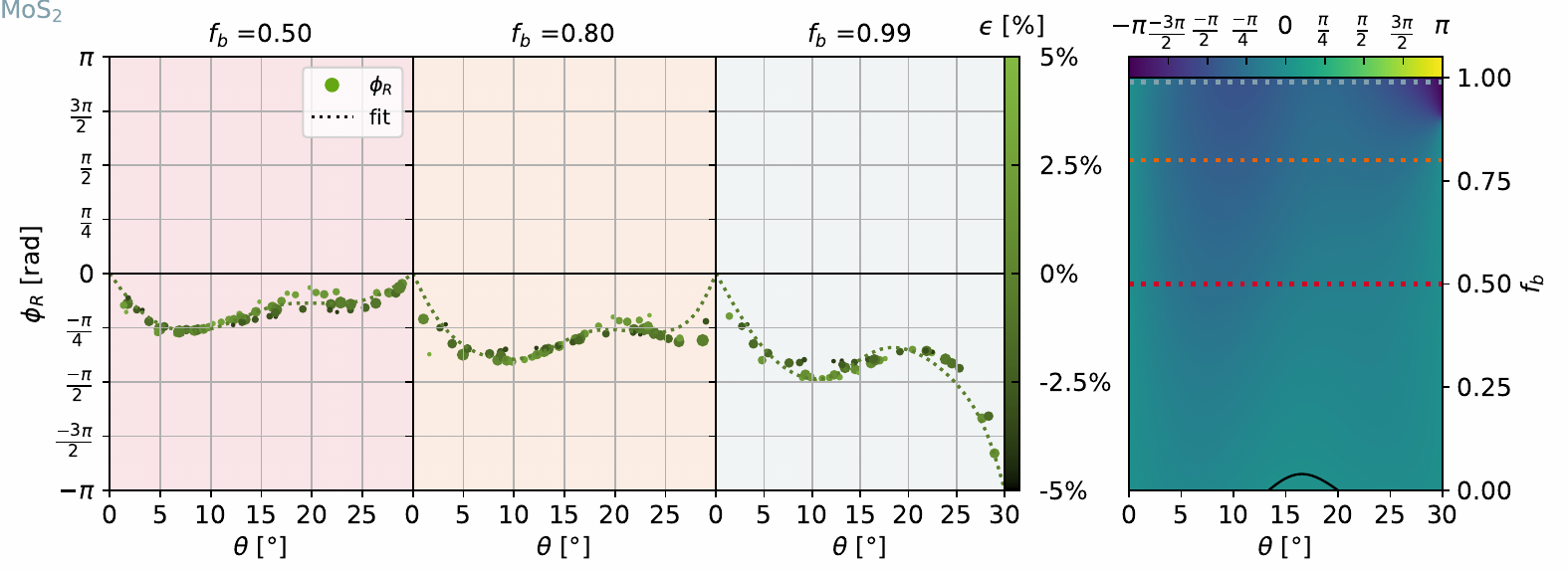}
                \caption{(left) $\rottot$-dependence of the Rashba phase of \mos\ at selected values of $\baf$.
                (right) Density plot of Rashba phase in the $(\rottot,\baf)$-parameter space.}
                \label{fig:apx:phir_mos2}
            \end{figure}
            
            The twist-angle dependence of the Rashba phase calculated for the band alignment values $\baf=\{0.50,0.80,0.99\}$ is shown in \figref{fig:apx:phir_mos2}, indicated by the dashed lines in the right panel.
            The range of variation of the Rashba phase is seen to increase as $\baf$ increases, with $\phi_R$ covering the full $[-\pi,0]$ range for $\baf=0.99$.
    \end{appendices}
    \clearpage
    \bibliography{References}
\end{document}